\pdfoutput=1
\documentclass[10pt,journal,compsoc]{IEEEtran}
\usepackage{times}
\usepackage{epsfig}
\usepackage{subfigure}
\usepackage{graphicx}
\usepackage{wrapfig}
\usepackage{amsmath}
\usepackage{amssymb}
\usepackage{multirow}
\usepackage{amsthm}
\usepackage{slashbox}

\bgroup\catcode`\#=12\egroup \newcommand{\tabincell}[2]{\begin{tabular}{@{}#1@{}}#2\end{tabular}}

\ifCLASSOPTIONcompsoc
\else
\fi
\ifCLASSINFOpdf
\else
\fi
\begin{document}
%
\title{A Comprehensive Survey on Cross-modal Retrieval}
%
%
%
%

\author{Kaiye Wang$^{\dagger}$,
        Qiyue Yin$^{\dagger}$,
        Wei Wang,
        Shu Wu,
        Liang Wang$^{*}$,~\IEEEmembership{Senior Member,~IEEE,}
\IEEEcompsocitemizethanks{\IEEEcompsocthanksitem Kaiye Wang, Qiyue Yin, Wei Wang, Shu Wu and Liang Wang are with the Center
for Research on Intelligent Perception and Computing (CRIPAC), National
Laboratory of Pattern Recognition (NLPR), Institute of Automation, Chinese
Academy of Sciences (CASIA), Beijing, 100190, China.\protect\\
E-mail: wangkaiye2005@126.com, \{qyyin,wangwei,shu.wu,wangliang\}@nlpr.
ia.ac.cn. }
\thanks{$^{\dagger}$ These two authors contribute equally to this work.}
\thanks{$^{*}$ The corresponding author.}}
\maketitle

\begin{abstract}
   In recent years, cross-modal retrieval has drawn much attention due to the rapid growth of multimodal data.
   It takes one type of data as the query to retrieve relevant data of another type.
   For example, a user can use a text to retrieve relevant pictures or videos.
   Since the query and its retrieved results can be of different modalities,
   how to measure the content similarity between different modalities of data remains a challenge.
   Various methods have been proposed to deal with such a problem.
   In this paper, we first review a number of representative methods for cross-modal retrieval
   and classify them into two main groups: 1) real-valued representation learning,
   and 2) binary representation learning.
   Real-valued representation learning methods aim to learn real-valued common representations for different modalities of data.
   To speed up the cross-modal retrieval, a number of binary representation learning methods are proposed to map different modalities of data
   into a common Hamming space.
   Then, we introduce several multimodal datasets in the community, and show the experimental results on two commonly used multimodal datasets.
   The comparison reveals the characteristic of different kinds of cross-modal retrieval methods,
   which is expected to benefit both practical applications and future research.
   Finally, we discuss open problems and future research directions.
\end{abstract}



%

\section{Introduction}

\begin{figure}
    \centering
    \includegraphics[width=0.5\textwidth]{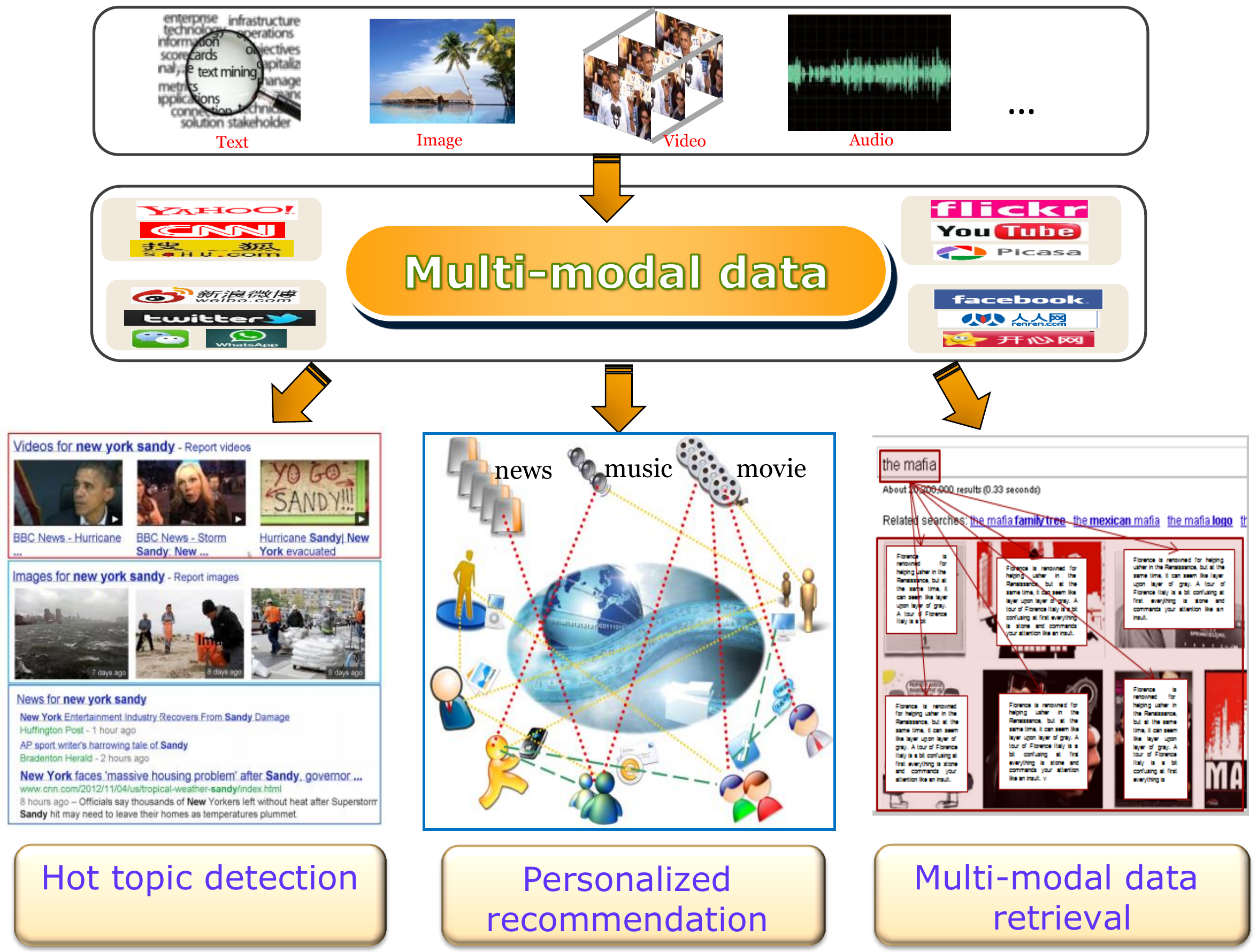}
    \caption{Multi-modal data and the promising applications.}
    \label{fig:multi-modalData}
\end{figure}

\IEEEPARstart{O}{ver} the last decade, different types of media data such as texts, images and videos are growing rapidly on the Internet.
It is common that different types of data are used for describing the same events or topics.
For example, a web page usually contains not only textual description but also images or videos for illustrating the common content.
Such different types of data are referred as \emph{multi-modal data}, which exhibit heterogeneous properties.
There have been many applications for multi-modal data (as shown in Figure \ref{fig:multi-modalData}).
As multimodal data grow, it becomes difficult for users to search information of interest effectively and efficiently.
Till now, there have been various research techniques for indexing and searching
multimedia data. However, these search techniques are mostly single-modality-based,
which can be divided into keyword-based retrieval and content-based retrieval.
They only perform similarity search of the same media type, such as text retrieval, image retrieval, audio retrieval, and video retrieval.
Hence, a demanding requirement for promoting information retrieval
is to develop a new retrieval model that can support the similarity search for multimodal data.

\begin{figure}
    \centering
    \includegraphics[width=0.5\textwidth]{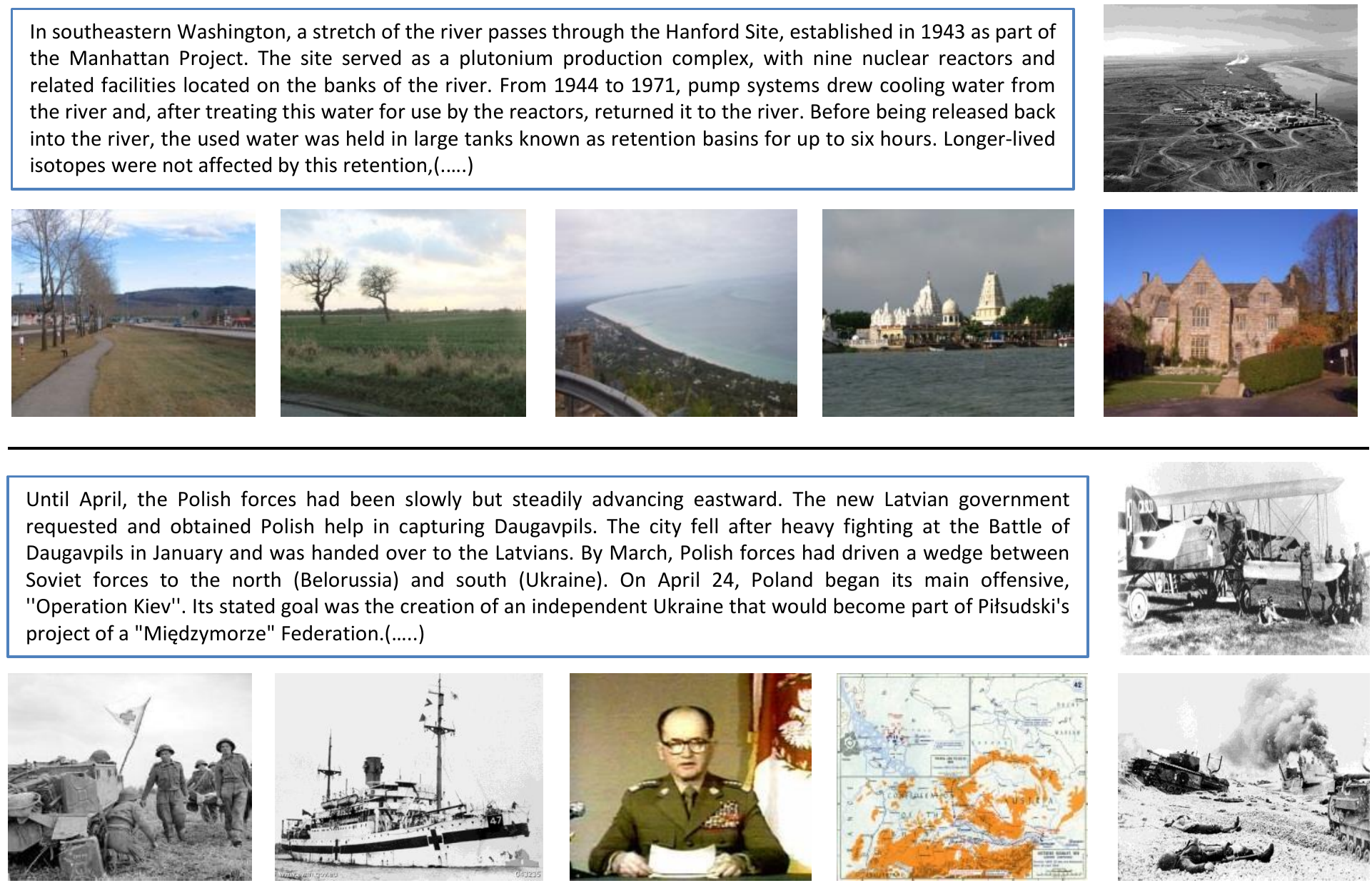}
    \caption{Two examples of cross-modal retrieval using text as query. The text query and its associated ground truth image are
    shown on the top, the retrieved images are shown at the bottom.}
    \label{fig:crossModalRetrieval}
\end{figure}

Nowadays, mobile devices and emerging social websites (e.g., Facebook, Flickr, YouTube, and
Twitter) are changing the ways people interact with the world and search
information of interest.
It is convenient if users can submit any media content at hand as the query.
Suppose we are on a visit to the Great Wall, by taking a photo, we may expect to use the photo
to retrieve the relevant textual materials as visual guides for us.
Therefore, cross-modal retrieval, as a natural searching way, becomes increasingly important.
Cross-modal retrieval aims to take one type of data as the query to retrieve relevant data
of another type.
For example, as shown in Figure \ref{fig:crossModalRetrieval}, the text is used as the query to retrieve images.
Furthermore, when users search information by submitting a query of any media type, they can
obtain search results across various modalities, which is more comprehensive
given that different modalities of data can provide complementary information to each other.

More recently, cross-modal retrieval has attracted considerable research attention.
The challenge of cross-modal retrieval is how to measure the content similarity between different modalities of data,
which is referred as the \emph{heterogeneity gap}.
Hence, compared with traditional retrieval methods, cross-modal retrieval requires cross-modal relationship modeling, so that
users can retrieve what they want by submitting what they have.
Now, the main research effort is to design the effective ways to
make the cross-modal retrieval more accurate and more scalable.

This paper aims to conduct a comprehensive survey of cross-modal retrieval.
Although Liu et al. \cite{Liu2014Survey} gave an overview of cross-modal retrieval in 2010,
it does not include many important works proposed in recent years.
Xu et al. \cite {Xu2013Survey} summarize several methods for modeling multimodal data, but they focus on multi-view learning.
Since many technical challenges remain in cross-modal retrieval, various ideas and techniques have been provided to solve the cross-modal problem in recent years.
This paper focuses on summarizing these latest works in cross-modal retrieval, the major concerns of which are very different from previous related surveys.
Another topic for modeling multimodal data is image/video description
\cite{Karpathy15MultimodalRNN,Oriol15CaptionGenerator,Chen15MindEye,Jia15LSTM,Yoshitaka15CoSMoS,mao2015deep,Kiros2015LanModel,Venugopalan2015translateVideo,Venugopalan2015videoTotext,Jeff15LSTM-CNN},
which is not discussed here because it goes beyond the scope of cross-modal retrieval research.

The major contributions of this paper are briefly summarized as follows.
\begin{itemize}
\item This paper aims to provide a survey on recent progress in cross-modal retrieval.
It contains many new references not found in previous surveys, which is
beneficial for the beginners to get familiar with cross-modal retrieval quickly.
\item This paper gives a taxonomy of cross-modal retrieval approaches.
Differences between different kinds of methods are elaborated, which are helpful for readers to better understand various techniques utilized
in cross-modal retrieval.
\item This paper evaluates several representative algorithms on the commonly used datasets. Some meaningful findings are obtained,
which are useful for understanding the cross-modal retrieval algorithms,
and are expected to benefit both practical applications and future research.
\item This paper summarizes challenges and opportunities in cross-modal retrieval fields, and points out some open directions in future.
\end{itemize}

The rest of this paper is organized as follows: we firstly give an overview on different kinds of methods for
cross-modal retrieval in Section 2.
Then, we illustrate different kinds of cross-modal retrieval algorithms in detail in Sections 3 and 4.
We introduce several multimodal datasets in Section 5.
Experimental results are reported in Section 6.
The discussion and future trends are given in Section 7.
Finally, Section 8 concludes this paper.

\section{Overview}
\label{sec:relatedWork}

\begin{figure}
    \centering
    \includegraphics[width=0.45\textwidth]{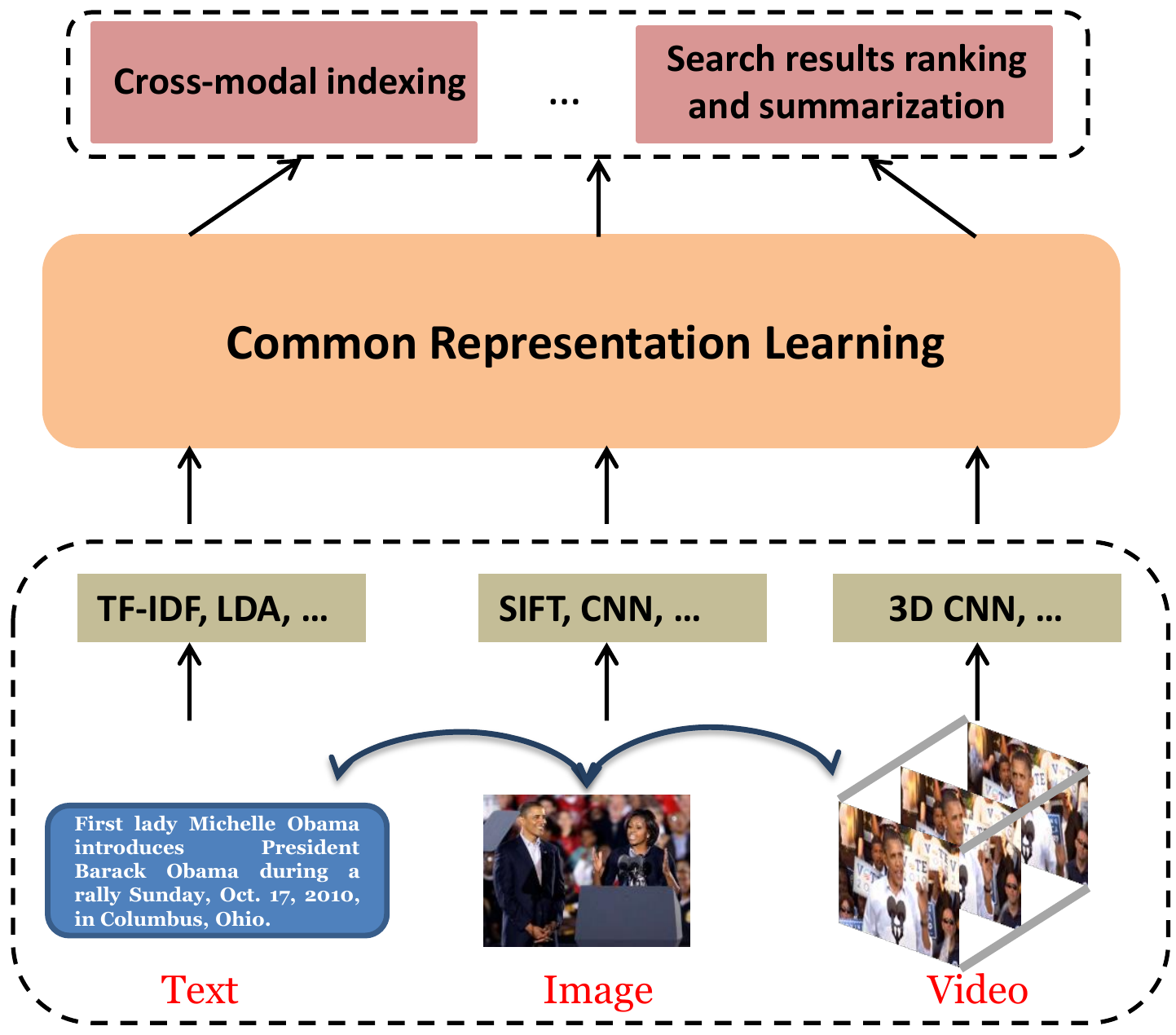}
    \caption{The general framework of cross-modal retrieval.}
    \label{fig:framework}
\end{figure}

In the cross-modal retrieval procedure, users can search various modalities of data
including texts, images and videos, starting with any modality of data as a query.
Figure \ref{fig:framework} presents the general framework of cross-modal retrieval,
in which, feature extraction for multimodal data is considered as the first step to
represent various modalities of data.
Based on these representations of multimodal data, cross-modal correlation modeling
is performed to learn common representations for various modalities of data.
At last, the common representations enable the cross-modal retrieval
by suitable solutions of search result ranking and summarization.

\begin{table*}
\caption{The summarization of typical algorithms for cross-modal retrieval.}
\begin{center}
\begin{tabular}{|c|c|c|c|}
\hline
\multicolumn{3}{|c|}{Category}      & Typical algorithms     \\

\hline\hline
\multirow{8}{*}{Real-valued representation learning}   & \multirow{4}{*}{Unsupervised methods}   & Subspace learning methods   &  \tabincell{c}{CCA \cite{ref06}, PLS \cite{ref18}, BLM \cite{ref32,ref30}, CFA \cite{ref51},  \\
MCU \cite{ref63}, CoCA \cite{shi2012dimensionality}, MMD \cite{zhu2014cross} }   \\
\cline{3-4}
                                                           &                &  Topic model   &    Corr-LDA \cite{ref48}, Tr-mm LDA  \cite{ref49}, MDRF \cite{ref50}   \\
\cline{3-4}
                                                            &                 &  Deep learning methods   &    \tabincell{c}{Multimodal Deep Autoencoder \cite{Ngiam2011MDL}, Multimodal DBM \cite{ref69},  \\
                                                           DCCA \cite{andrew2013deep}, End-to-end DCCA \cite{Yan2015DCCA}, Corr-AE \cite{feng2014cross}, \\
                                                           Joint Video-Language Model \cite{xu2015jointly} }  \\
\cline{2-4}
                                 & \multirow{3}{*}{Pairwise based methods}       &  Shallow methods   &  Multi-NPP \cite{ref38}, MVML-GL \cite{zhai2012multiview}, JGRHML \cite{ref67} \\
\cline{3-4}
                                                           &                                  &  Deep learning methods   &   RGDBN \cite{yuan2013latent}, MSDS \cite{Wang15ImageText}  \\
\cline{2-4}
                                    & \multirow{3}{*}{Rank based methods}       &  Shallow methods   &  \tabincell{c}{SSI \cite{bai2010learning}, PAMIR \cite{grangier2008discriminative}, LSCMR \cite{ref55}, bi-LSCMR \cite{ref56}, \\
                                                            Wsabie \cite{Weston2011Wsabie}, RCCA \cite{Yao15RCCA}  } \\
\cline{3-4}
                                                           &                         &  Deep learning methods   &  \tabincell{c}{DeViSE \cite{frome2013devise}, DT-RNNs \cite{socher2014grounded}, Deep Fragment \cite{Karpathy14DeepFragment}, \\
                                                           C$^2$MLR \cite{Jiang2015C2MLR} }   \\
\cline{2-4}
                                      & \multirow{3}{*}{Supervised methods}       &  Subspace learning methods   & \tabincell{c}{CDFE \cite{ref19}, GMA \cite{ref32}, I$^2$SCA \cite{jing2014intra}, PFAR \cite{ref64}, JRL \cite{zhai2014learning},  \\
                       SliM$^2$ \cite{zhuang2013supervised}, cluster-CCA \cite{rasiwasia2014cluster}, CCA-3V \cite{gong2014multi}, LCFS \cite{ref52},   \\
                        JFSSL \cite{Wang2016JFSSL}  }  \\
\cline{3-4}
                                                           &      &  Topic model   &   SupDocNADE \cite{zheng2014topic}, NPBUS \cite{liao2014nonparametric}, M$^3$R \cite{wang2014multi}  \\
\cline{3-4}
                                                           &      &  Deep learning methods   & RE-DNN \cite{Wang15DeepMapping}, deep-SM \cite{Wei16NewBaseline}, MDNN \cite{Wang16EffectiveDeep}    \\

\hline
\multirow{8}{*}{Binary representation learning}   & \multirow{3}{*}{Unsupervised methods}   & Linear modeling   &   \tabincell{c}{CVH \cite{ref14}, IMH \cite{song2013inter}, PDH \cite{Rastegari2013PDH}, LCMH \cite{Zhu2013LCMH}, CMFH \cite{ding2014collective}, \\
                        LSSH \cite{zhou2014latent}  }      \\
\cline{3-4}
                                                           &                    &  Nonlinear modeling   &   MSAE \cite{wang2014effective}, DMHOR \cite{Wang15DMHOR}   \\
\cline{2-4}
                                                    & \multirow{2}{*}{Pairwise based methods}       &  Linear modeling   &   \tabincell{c}{CMSSH \cite{ref13}, CRH \cite{zhen2012co}, IMVH \cite{hu2014iterative},  QCH\cite{Wu2015QCH} \\
                         RaHH \cite{ou2013comparing}, HTH \cite{wei2014scalable} }  \\
\cline{3-4}
                                                           &       &  Nonlinear modeling   & MLBE \cite{Zhen2012MLBE}, PLMH \cite{zhai2013parametric},  MM-NN \cite{masci2014multimodal}, CHN \cite{Cao16CHN}  \\
\cline{2-4}
                                                           & \multirow{2}{*}{Supervised methods}       &  Linear modeling   &  SM$^2$H \cite{wu2014sparse}, DCDH \cite{yu2014discriminative}, SCM \cite{zhang2014large}   \\
\cline{3-4}
                                                           &                 &  Nonlinear modeling   &   SePH \cite{Lin15SePH}, CAH \cite{Cao16CAH}, DCMH \cite{JiangL16DeepCrossHash}  \\

\hline
\end{tabular}
\end{center}

\label{tab:summarization}
\end{table*}

Since the cross-modal retrieval is considered as an important problem in real applications,
various approaches have been proposed to deal with this problem,
which can be roughly divided into two categories: 1) real-valued representation learning and
2) binary representation learning, which is also called cross-modal hashing.
For real-valued representation learning, the learned common representations for various modalities of data are real-valued.
To speed up cross-modal retrieval, the binary representation learning methods aim to transform different modalities of data into a common
Hamming space, in which cross-modal similarity search is fast.
Since the representation is encoded to binary codes, the retrieval accuracy generally decreases slightly due to the loss of information.


According to the utilized information when learning the common representations,
the cross-modal retrieval methods can be further divided into four groups:
1) unsupervised methods, 2) pairwise based methods, 3) rank based methods, and 4) supervised methods.
Generally speaking, the more information one method utilizes, the better performance it obtains.

1) For unsupervised methods, only co-occurrence information is utilized to learn common representations across multi-modal data.
The co-occurrence information means that if different modalities of data are co-existed in a multimodal document,
then they are of the same semantic. For example, a web page usually contains both
textual descriptions and images for illustrating the same event or topic.

2) For the pairwise based methods, similar pairs (or dissimilar pairs) are utilized to learn common representations.
These methods generally learn a meaningful metric distance between different modalities of data.

3) For the rank based methods, rank lists are often utilized to learn common representations.
Ranking based methods study the cross-modal retrieval as a problem of learning to rank.

4) Supervised methods exploit label information to learn common representations.
These methods enforce the learned representations of different-class samples to be far apart while those of the same-class samples lie as close as possible.
Accordingly, they obtain more discriminative representations.
But getting label information is sometimes expensive due to massive manual annotation.

%
%
%

Typical algorithms of the cross-modal retrieval in terms of different categories are summarized in Table \ref{tab:summarization}.

\section{Real-valued Representation Learning}

If different modalities of data are related to the same event or topic,
they are expected to share certain common representation space in which relevant
data are close to each other. Real-valued representation learning methods
aim to learn a real-valued common representation space, in which different modalities of data
can be directly measured.
According to the information utilized to learn the common representation,
the cross-modal retrieval methods can be further divided into four groups:
1) unsupervised methods, 2) pairwise based methods, 3) rank based methods, and 4) supervised methods.
We will introduce them in the following, respectively,
and describe some of them in details for better understanding.

\subsection{Unsupervised methods}

The unsupervised methods only utilize co-occurrence information to learn common representations across multi-modal data.
The co-occurrence information means that if different modalities of data are co-existed in a multimodal document,
then they are of the similar semantic. For example,
the textual description along with images or videos often exist in a webpage to illustrate the same event or topic.
Furthermore, the unsupervised methods are categorized into subspace learning methods, topic models and deep learning methods.

\subsubsection{Subspace learning methods}

\begin{figure}
    \centering
    \includegraphics[width=0.45\textwidth]{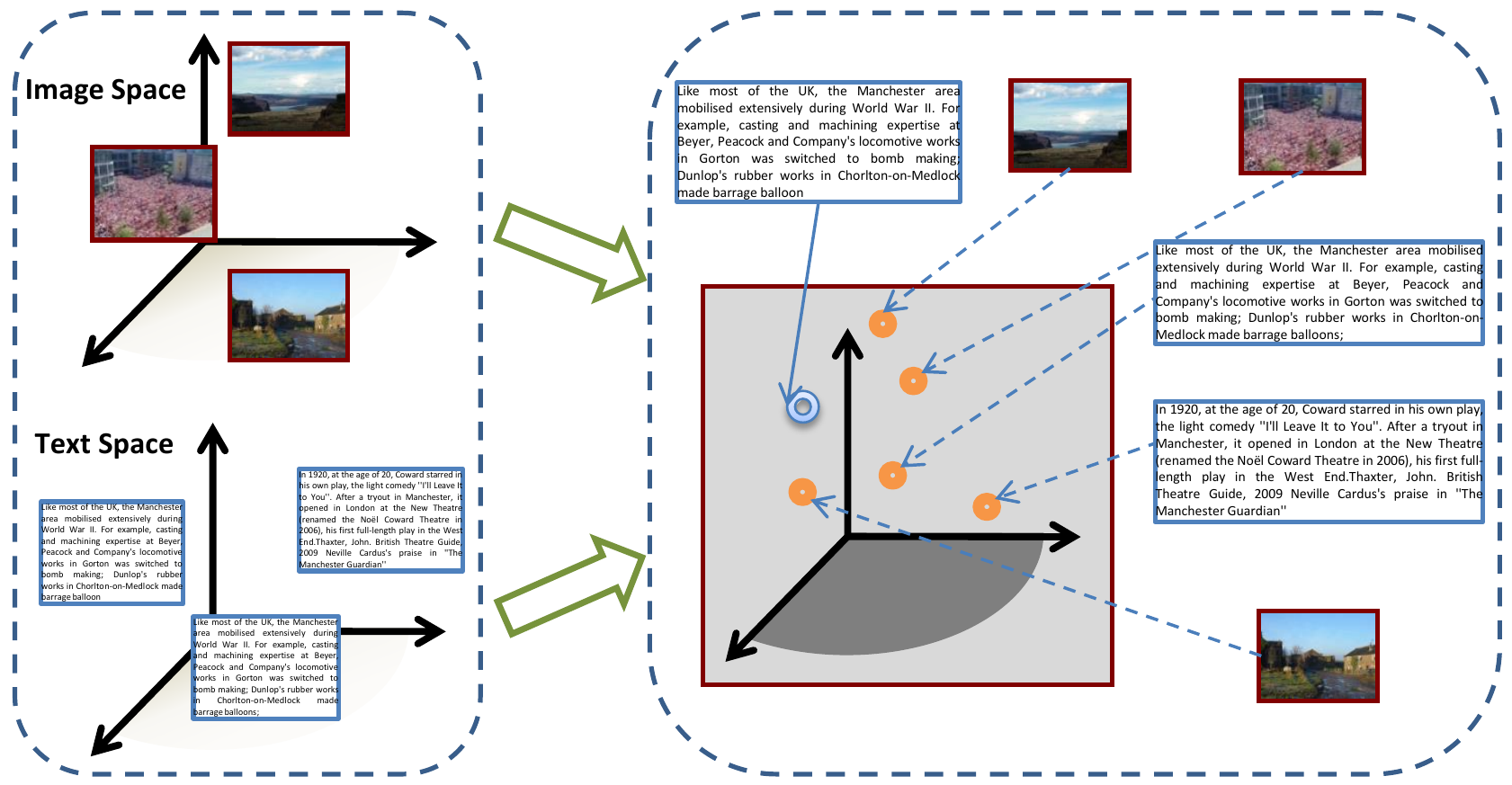}
    \caption{The sketch map of subspace learning methods for multimodal data (take images and texts as example).}
    \label{fig:subspace}
\end{figure}

The main difficulty of cross-modal retrieval is how to measure the content similarity between different modalities of data.
Subspace learning methods are one type of the most popular methods. They aim to learn a common subspace shared by
different modalities of data, in which the similarity between different modalities of data can be measured (as shown in Figure \ref{fig:subspace}).
Unsupervised subspace learning methods use pairwise information to learn a common latent subspace across multi-modal data.
They enforce pair-wise closeness between different modalities of data in the common subspace.

Canonical Correlation Analysis (CCA) is one of the most popular unsupervised subspace learning methods
for establishing inter-modal relationships between data from different modalities.
It has been widely used for cross-media retrieval\cite{ref06,ref61,ref62}, cross-lingual retrieval \cite{ref31} and some vision problems \cite{ref04}.
CCA aims to learn two directions ${{\bf{w}}_x}$ and ${{\bf{w}}_y}$ for two modalities of data$(x,y)$, along which
the data is maximally correlated, i.e.,
\begin{equation}\label{eq:cca}
\mathop {\max }\limits_{{{\bf{w}}_x},{{\bf{w}}_y}} \frac{{{\bf{w}}_x^T{\Sigma_{xy}}{{\bf{w}}_y}}}
{{\sqrt {{\bf{w}}_x^T{\Sigma_{xx}}{{\bf{w}}_x}} \sqrt {{\bf{w}}_y^T{\Sigma_{yy}}{{\bf{w}}_y}} }}
\end{equation}
where ${{\Sigma_{xx}}}$ and ${{\Sigma_{yy}}}$ represent the empirical covariance matrices for the two modalities of data respectively,
while ${\Sigma_{xy}} = \Sigma_{yx}^T$ represents the cross-covariance matrix between them.
Rasiwasia et al.~\cite{ref06} propose a two-stage method for cross-modal multimedia retrieval.
In the first stage, CCA is used to learn a common subspace by maximizing the correlation between the two modalities.
Then, a semantic space is learned to measure the similarity of different modal features.

Besides CCA, Partial Least Squares (PLS) \cite{ref18} and Bilinear Model (BLM) \cite{ref32,ref30}
are also used for cross-modal retrieval.
Sharma and Jacobs \cite{ref05} use PLS to linearly map images with different modalities
to a common linear subspace in which they are highly correlated.
Chen et al.~\cite{ref07} apply PLS to the cross-modal document retrieval.
They use PLS to switch the image features into the text space, and then learn a semantic space
for the measure of similarity between two different modalities.
In \cite{ref30}, Tenenbaum and Freeman propose a bilinear model (BLM) to derive a common space for cross-modal face recognition.
BLM is also used for text-image retrieval in \cite{ref32}.

Li et al. \cite{ref51} introduce a cross-modal factor analysis (CFA) approach to evaluate the association between two modalities.
The CFA method adopts a criterion of minimizing the Frobenius norm between pairwise data in the transformed domain.
Mahadevan et al. \cite{ref63} propose maximum covariance unfolding (MCU), a manifold learning algorithm for
simultaneous dimensionality reduction of data from different modalities.
Shi et al. \cite{shi2012dimensionality} propose a principle of collective component analysis (CoCA),
to handle dimensionality reduction on a heterogeneous feature space.
Zhu et al. \cite{zhu2014cross} propose a greedy dictionary construction method for the cross-modal retrieval problem.
The compactness and modality-adaptivity are preserved by including reconstruction error terms and a Maximum Mean Discrepancy (MMD)
measurement for both modalities in the objective function.
Wang et al. \cite{wang2013cross} propose to learn the sparse projection matrices that map the image-text pairs in Wikipedia into
a latent space for cross-modal retrieval.

\subsubsection{Topic models}

Another unsupervised method is the topic model. Topic models have been widely applied to a specific cross-modal problem, i.e., image annotation \cite{ref48,ref49}.
To capture the correlation between images and annotations, Latent Dirichlet Allocation (LDA) \cite{ref42} has been extended to
learn the joint distribution of multi-modal data, such as Correspondence LDA (Corr-LDA) \cite{ref48} and Topic-regression Multi-modal LDA (Tr-mm LDA) \cite{ref49}.
Corr-LDA uses topics as the shared latent variables, which represent the underlying causes of cross-correlations in the multi-modal data.
Tr-mm LDA learns two separate sets of hidden topics and a regression module
which captures more general forms of association and allows one set of topics to be linearly predicted from the other.

Jia et al. \cite{ref50} propose a new probabilistic model (Multi-modal Document Random Field, MDRF) to learn a set of shared topics across the modalities. The model defines
a Markov random field on the document level which allows modeling more flexible document similarities.

\subsubsection{Deep learning methods}

\begin{figure}
    \centering
    \includegraphics[width=0.45\textwidth]{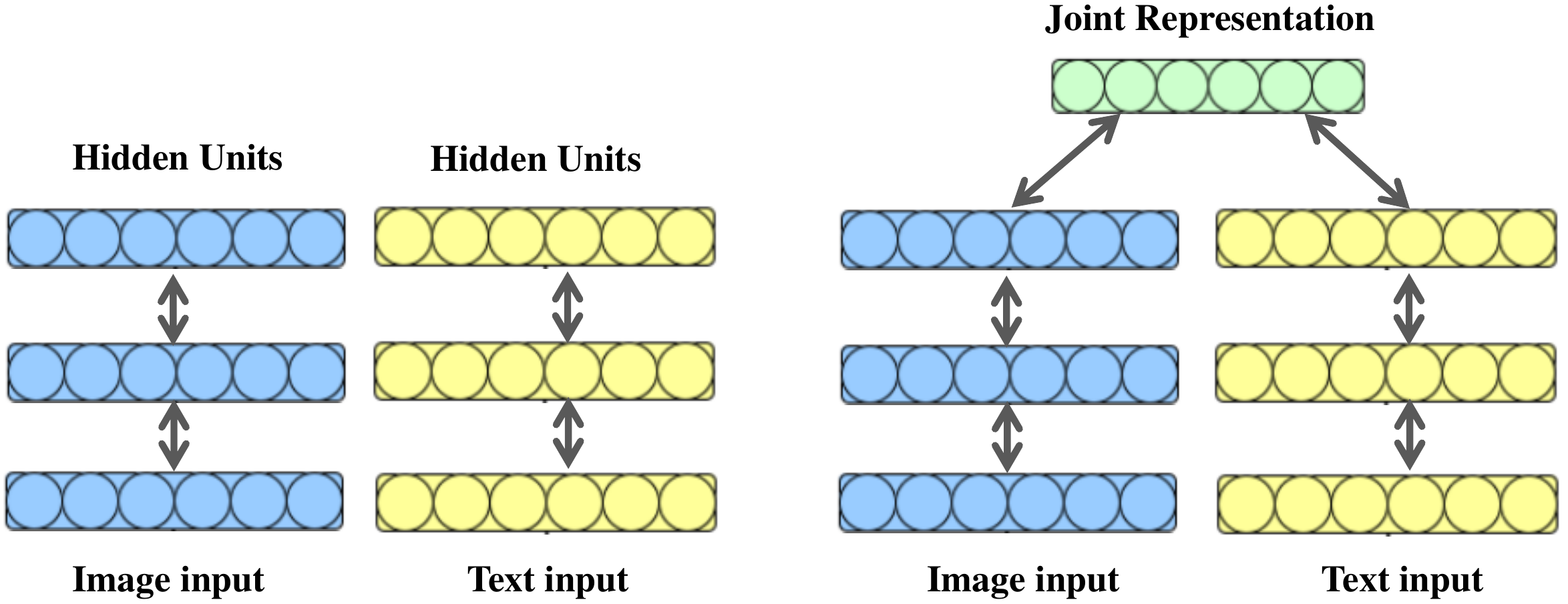}
    \caption{Learn the joint representations for multimodal data (e.g., images and texts) using deep learning model ( adapted from \cite{ref69}).}
    \label{fig:DBM}
\end{figure}

As we mentioned above, it is common that different types of data are used for description of the same events or topics in the web.
For example, user-generated content usually involves with data from different modalities, such as images, texts and videos.
This makes it very challenging for traditional methods to obtain a joint representation for multimodal data.
Inspired by recent progress of deep learning, Ngiam et al. \cite{Ngiam2011MDL} apply deep networks to learn features over multiple modalities, which
focuses on learning representations for speech audio that are coupled with videos of the lips.
Then, a deep Restricted Boltzmann Machine \cite{ref69} succeeds in
learning the joint representations for multimodal data.
It firstly uses separate modality-friendly latent models to learn low-level representations for each modality, and
then fuses into joint representation along the deep architecture in the higher-level (as shown in Figure \ref{fig:DBM}).

\begin{figure*}
    \centering
    \includegraphics[width=0.8\textwidth]{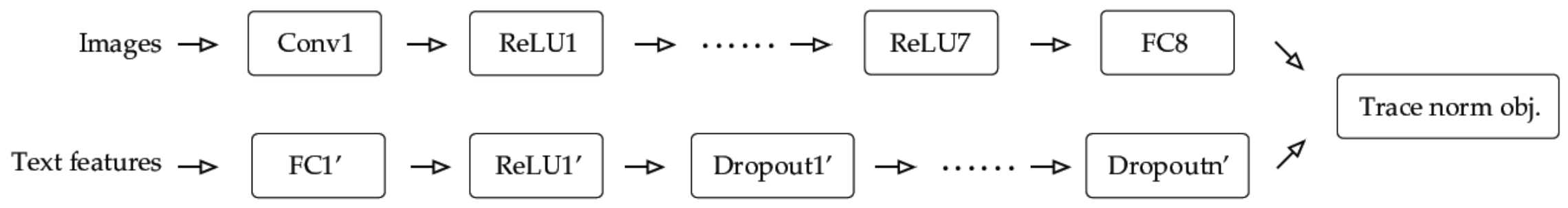}
    \caption{Architecture of end-to-end deep canonical correlation analysis (adapted from \cite{Yan2015DCCA}).}
    \label{fig:DeepCCA}
\end{figure*}

Inspired by representation learning using deep networks \cite{Ngiam2011MDL,ref69}, Andrew et al. \cite{andrew2013deep} present
Deep Canonical Correlation Analysis (DCCA), a deep learning method to learn complex nonlinear projection for
different modalities of data such that the resulting representations are highly linearly correlated.
Furthermore, Yan and Mikolajczyk \cite{Yan2015DCCA} propose an end-to-end learning
scheme based on the deep canonical correlation analysis (End-to-end DCCA),
which is a non-trivial extension to \cite{andrew2013deep} (as shown in Figure \ref{fig:DeepCCA}).
The objective function is
\begin{equation}
\begin{array}{l}
 \mathop {\max }\limits_{{{\bf{W}}_x},{{\bf{W}}_y}} {\rm{ }}tr({\bf{W}}_x^T{\Sigma_{xy}}{{\bf{W}}_y}) \\
 s.t.{\rm{ }}:{\bf{W}}_x^T{\Sigma_{xx}}{{\bf{W}}_x} = {\bf{W}}_y^T{\Sigma_{yy}}{{\bf{W}}_y} = I \\
 \end{array}
\end{equation}
Define ${\bf{T}} = \Sigma_{xx}^{ - 1/2}{\Sigma_{xy}}\Sigma_{yy}^{ - 1/2}$, then the objective function is rewritten as
\begin{equation}
corr({\bf{X}},{\bf{Y}}) = tr({({{\bf{T}}^{\bf{T}}}{\bf{T}})^{{1 \mathord{\left/
 {\vphantom {1 2}} \right.
 \kern-\nulldelimiterspace} 2}}})
\end{equation}
The gradients with respect to ${\bf{X}}$ and ${\bf{Y}}$ are computed and propagated down along the two branches of the network.
The high dimensionality of features presents a great challenge in terms of memory
and speed complexity when used in the DCCA framework.
To address this problem, Yan and Mikolajczyk propose and discuss details of a GPU implementation
with CULA libraries. The efficiency of the implementation
is several orders of magnitude higher than CPU implementation.

Feng et al. \cite{feng2014cross} propose a novel model involving correspondence autoencoder
(Corr-AE) for cross-modal retrieval. The model is constructed by correlating hidden representations of two uni-modal autoencoders.
A novel objective, which minimizes a linear combination of representation
learning errors for each modality and correlation learning errors between hidden representations of two modalities,
is utilized to train the model as a whole.
Minimization of correlation learning errors forces the model to learn hidden representations
with only common information in different modalities,
while minimization of representation learning errors makes
hidden representations good enough to reconstruct the input of each modality.

Xu et al. \cite{xu2015jointly} propose a unified framework that jointly models video and the corresponding text sentences.
The framework consists of three parts: a compositional semantics language model, a deep video model and a
joint embedding model. In their language model, they propose a
dependency-tree structure model that embeds sentences into a
continuous vector space, which preserves visually grounded
meanings and word order. In the visual model, they leverage
deep neural networks to capture essential semantic information from videos.
In the joint embedding model, they minimize the distance of the outputs of the deep video model and
compositional language model in the joint space, and update these two models jointly.
Based on these three parts, this model
is able to accomplish three tasks:
1) natural language generation, 2) video-language retrieval and 3) language-video retrieval.

\subsection{Pairwise based methods}

\begin{figure*}
    \centering
    \includegraphics[width=0.8\textwidth]{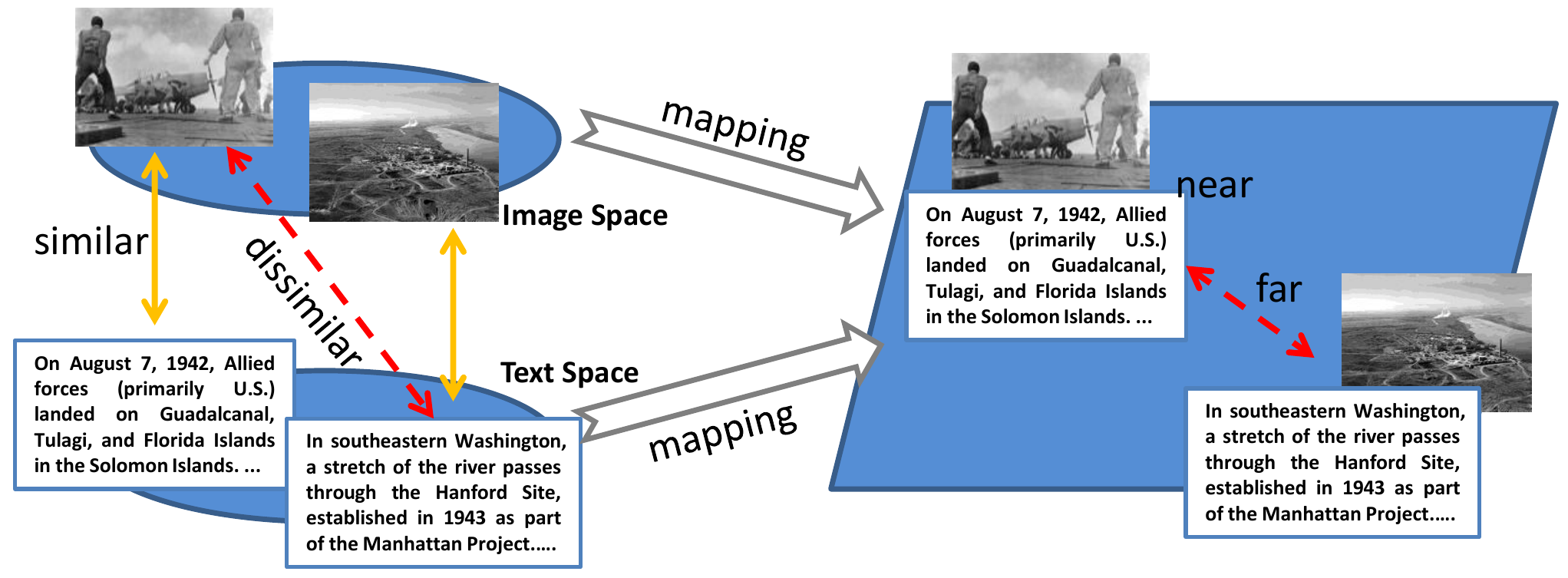}
    \caption{The basic idea of heterogeneous metric learning, which learns projections using similar/dissimilar pairs.}
    \label{fig:MericLearning}
\end{figure*}

Compared with the unsupervised method, pairwise based methods utilize more similar pairs (or dissimilar pairs)
to learn a meaningful metric distance between different modalities of data,
which can be regarded as heterogeneous metric learning (as shown in Figure \ref{fig:MericLearning}).

\subsubsection{Shallow methods}

Wu et al. \cite{ref39} study the metric learning problem to find a similarity function over two different spaces.
Mignon and Jurie \cite{ref20} propose a metric learning approach for cross-modal matching, which considers both positive
and negative constraints.
Quadrianto and Lampert \cite{ref38} propose a new metric learning scheme (Multi-View Neighborhood Preserving
Projection, Multi-NPP) to project different modalities into
a shared feature space, in which the Euclidean distance provides a meaningful intra-modality and inter-modality similarity.
To learn projections $g_1^{{w_1}}({{\bf{x}}_i})$ and $g_2^{{w_2}}({{\bf{y}}_j})$ for different features ${\bf{x}}_i$ and ${\bf{y}}_j$, the loss function is defined as

\begin{equation}\label{eq:multi-npp}
\begin{array}{l}
 {L^{i,j}}({{\bf{w}}_1},{{\bf{w}}_2},{{\bf{x}}_i},{{\bf{y}}_j},{S_{{x_i}}}) =  \\
 \frac{{{{\bf{I}}_{\left[\kern-0.15em\left[ {{y_j} \in {S_{{x_i}}}}
 \right]\kern-0.15em\right]}}}}{2} \times L_1^{i,j} + \frac{{\left( {1 - {{\bf{I}}_{\left[\kern-0.15em\left[ {{y_j} \in {S_{{x_i}}}}
 \right]\kern-0.15em\right]}}} \right)}}{2} \times L_2^{i,j} \\
 \end{array}
\end{equation}
with
\begin{equation}\label{eq:multi-npp2}
L_1^{i,j} = \left\| {g_1^{{w_1}}({{\bf{x}}_i}) - g_2^{{w_2}}({{\bf{y}}_j})} \right\|_{Fro}^2
\end{equation}
\begin{equation}\label{eq:multi-npp3}
L_2^{i,j}({\beta _d}) = \left\{ \begin{array}{l}
  - \frac{1}{2}\beta _d^2 + \frac{{a{\lambda ^2}}}{2},{\rm{    ~~~~~~~~~             }}if~0 \le \left| {{\beta _d}} \right| < \lambda  \\
 \frac{{{{\left| {{\beta _d}} \right|}^2} - 2a\lambda \left| {{\beta _d}} \right| + {a^2}{\lambda ^2}}}{{2(a - 1)}},{\rm{  ~~}}if~\lambda  \le \left| {{\beta _d}} \right| \le a\lambda  \\
 0,{\rm{           ~~~~~~~~~~~~~~~~~~~~~~                         }}if~\left| {{\beta _d}} \right| \ge a\lambda  \\
 \end{array} \right.
\end{equation}
where ${\beta _d} = {\left\| {g_1^{{w_1}}({{\bf{x}}_i}) - g_2^{{w_2}}({{\bf{y}}_j})} \right\|_{Fro}}$ for appropriately
chosen constants $a$ and $\lambda$.
The above loss function consists of the similarity term $L_1^{i,j}$ that enforces similar
objects to be at proximal locations in the latent space and the dissimilarity term $L_2^{i,j}$ that pushes
dissimilar objects away from each other.

Zhai et al. \cite{zhai2012multiview} propose a new method called Multiview Metric Learning with Global consistency and Local smoothness (MVML-GL).
This framework consists of two main
steps. In the first step, they seek a global consistent shared latent feature space.
In the second step, the explicit mapping functions between the input spaces and the shared
latent space are learned via regularized local linear regression.
Zhai et al. \cite{ref67} propose a joint graph regularized heterogeneous metric learning (JGRHML) algorithm to learn a heterogeneous metric for cross-modal retrieval.
Based on the heterogeneous metric, they further learn a high-level semantic metric through label propagation.

\subsubsection{Deep learning methods}

To predict the links between social media,
Yuan et al. \cite{yuan2013latent} design a Relational Generative Deep Belief Nets (RGDBN) model to learn latent features
for social media, which utilizes the relationships between social media in the network.
In the RGDBN model, the link between items is generated from the interactions of their latent features.
By integrating the Indian buffet process into the modified Deep Belief Nets, they learn
the latent feature that best embeds both the media content and observed media relationships.
The model is able to analyze the links between heterogeneous as well as homogeneous data,
which can also be used for cross-modal retrieval.

Wang et al. \cite{Wang15ImageText} propose a novel model based on
modality-specific feature learning, named as Modality-Specific Deep Structure (MSDS).
Considering the characteristics of different modalities, the model uses two
types of convolutional neural networks to map the raw data
to the latent space representations for images and texts, respectively.
Particularly, the convolution based network used
for texts involves word embedding learning, which has been
proved effective to extract meaningful textual features for
text classification. In the latent space, the mapped features
of images and texts form relevant and irrelevant image-text
pairs, which are used by the one-vs-more learning scheme.

\subsection{Rank based methods}

The rank based methods utilize rank lists to learn common representations.
Rank based methods study the cross-modal retrieval as a problem of learning to rank.

\subsubsection{Shallow methods}

Bai et al. \cite{bai2010learning} present Supervised Semantic Indexing (SSI) for cross-lingual retrieval.
Grangier et al. \cite{grangier2008discriminative} propose a discriminative kernel-based method (Passive-Aggressive Model for Image Retrieval, PAMIR)
to solve the problem of cross-modal ranking by adapting the Passive-Aggressive algorithm.
Weston et al. \cite{Weston2011Wsabie} introduce a scalable model for image annotation by learning a joint representation of images
and annotations. It learns to optimize precision at the top of the ranked list of annotations for a given image and
learns a low-dimensional joint embedding space for both images and annotations.

Lu et al. \cite{ref55} propose a cross-modal ranking algorithm for cross-modal retrieval, called Latent Semantic Cross-Modal Ranking (LSCMR).
They utilize the structural SVM to learn a metric such that ranking of data induced by the distance from a query can be optimized against
various ranking measures.
However, LSCMR does not make full use of bi-directional ranking examples (bi-directional
ranking means that both text-query-image and image-query-text ranking examples are utilized in the training).
Accordingly, Wu et al. \cite{ref56} propose to optimize the bi-directional listwise ranking loss with a latent space embedding.

Recently, Yao et al. \cite{Yao15RCCA} propose a novel Ranking Canonical
Correlation Analysis (RCCA) for learning query and image similarities.
RCCA is used to adjust the subspace learnt by CCA to further preserve the preference relations in the click data.
The objective function of the RCCA is
\begin{equation}\label{eq:RCCA}
\begin{array}{l}
 \mathop {\arg \min }\limits_{{{\bf{W}}_q},{{\bf{W}}_v},{\bf{W}}} L(s(q,v,{{\bf{W}}_q},{{\bf{W}}_v},{\bf{W}})) + \frac{\mu }{2}{\left\| {\bf{W}} \right\|^2} \\
 {\rm{ ~~~~~~~ }} + \left( {\frac{\gamma }{2}{{\left\| {{{\bf{W}}_q} - {\bf{W}}_q^0} \right\|}^2} + \frac{\eta }{2}{{\left\| {{{\bf{W}}_v} - {\bf{W}}_v^0} \right\|}^2}} \right) \\
 \end{array}
\end{equation}
where ${\bf{W}}_q^0$ and ${\bf{W}}_v^0$ are the initial transformation matrices learnt by CCA,
and $L$ is the margin ranking loss as follows:
\begin{equation}\label{eq:RCCA1}
\begin{array}{l}
 L(s(q,v,{{\bf{W}}_q},{{\bf{W}}_v},{\bf{W}})) =  \\
 \sum\limits_T {\max (0,1 - s(q,{v^ + },{{\bf{W}}_q},{{\bf{W}}_v},{\bf{W}}) + s(q,{v^ - },{{\bf{W}}_q},{{\bf{W}}_v},{\bf{W}}))}  \\
 \end{array}
\end{equation}
where $s(q,v,{{\bf{W}}_q},{{\bf{W}}_v},{\bf{W}})$ is a query-image similarity function
that is used to measure the similarity of image $v$ given query $q$ in the latent space.
\begin{equation}\label{eq:RCCA2}
s(q,v,{{\bf{W}}_q},{{\bf{W}}_v},{\bf{W}}) = (q{{\bf{W}}_q}){\bf{W}}{(v{{\bf{W}}_v})^T}
\end{equation}

\subsubsection{Deep learning methods}

Inspired by the progress of deep learning, Frome et al. \cite{frome2013devise} present a new deep visual-semantic embedding model (DeViSE), the objective of which
is to leverage semantic knowledge learned in the text domain, and transfer it to a model trained for visual object recognition.

Socher et al. \cite{socher2014grounded} introduce a Dependency Tree Recursive Neural Networks (DT-RNNs) which uses dependency
trees to embed sentences into a vector space in order to retrieve images that are described by those sentences.
The image features ${{\bf{z}}_i}$ are extracted from a deep neural network.
To learn joint image-sentence representations, the ranking cost function is:
\begin{equation}\label{eq:DT-RNN}
\begin{array}{l}
 J({{\bf{W}}_I},\theta ) = \sum\limits_{(i,j) \in P} {\sum\limits_{c \in S\backslash S(i)} {\max (0,\Delta  - {\bf{v}}_i^T{{\bf{y}}_j} + {\bf{v}}_i^T{{\bf{y}}_c})} }  \\
 {\rm{             ~~~~~~~~~~~~~  }} + \sum\limits_{(i,j) \in P} {\sum\limits_{c \in I\backslash I(j)} {\max (0,\Delta  - {\bf{v}}_i^T{{\bf{y}}_j} + {\bf{v}}_c^T{{\bf{y}}_j})} }  \\
 \end{array}
\end{equation}
where ${{\bf{v}}_i} = {{\bf{W}}_I}{{\bf{z}}_i}$ is the mapped image vector, and
${y_{ij}} = DTRN{N_\theta }({s_{ij}})$ is the composed sentence vector.
$s_{ij}$ is the $j$-th sentence description for image ${{\bf{z}}_i}$.
$S$ is the set of all sentence indices and $S_{i}$ is the set of sentence indices corresponding to image $i$.
Similarly, $I$ is the set of all image indices and $I_{j}$ is the set of image indices corresponding to sentence $j$.
$P$ is the set of all correct image-sentence training pairs $(i,j)$. With both images and sentences in the same
multimodal space, the image-sentence retrieval is performed easily.

Karpathy et al. \cite{Karpathy14DeepFragment} introduce a model for the bidirectional retrieval of images and sentences,
which formulates a structured, max-margin objective for a deep neural network
that learns to embed both visual and language data into a common, multimodal space.
Unlike previous models that directly map images or sentences into a common embedding space,
this model works on a finer level and embeds fragments of images (objects) and fragments of sentences (typed dependency tree relations) into a common space.

Jiang et al. \cite{Jiang2015C2MLR} exploit the existing image-text databases to optimize a ranking function for cross-modal retrieval,
called deep compositional cross-modal learning to rank (C$^2$MLR).
C$^2$MLR considers learning a multi-modal embedding from the perspective of optimizing a pairwise ranking problem
while enhancing both local alignment and global alignment. In particular, the local alignment
(i.e., the alignment of visual objects and textual words) and the global alignment (i.e., the image-level and sentence-level alignment)
are collaboratively utilized to learn the multi-modal common embedding space in a max-margin learning to rank manner.

Hua et al. \cite{Hua15CrossLearning} develop a novel deep convolutional
architecture for cross-modal retrieval, named Cross-Modal Correlation
learning with Deep Convolutional Architecture (CMCDCA).
It consists of visual feature representation learning and cross-modal correlation
learning with a large margin principle.

\subsection{Supervised methods}

To obtain a more discriminative common representation, supervised methods exploit label information,
which provides a much better separation between classes in the common representation space.

\begin{figure}
    \centering
    \includegraphics[width=0.45\textwidth]{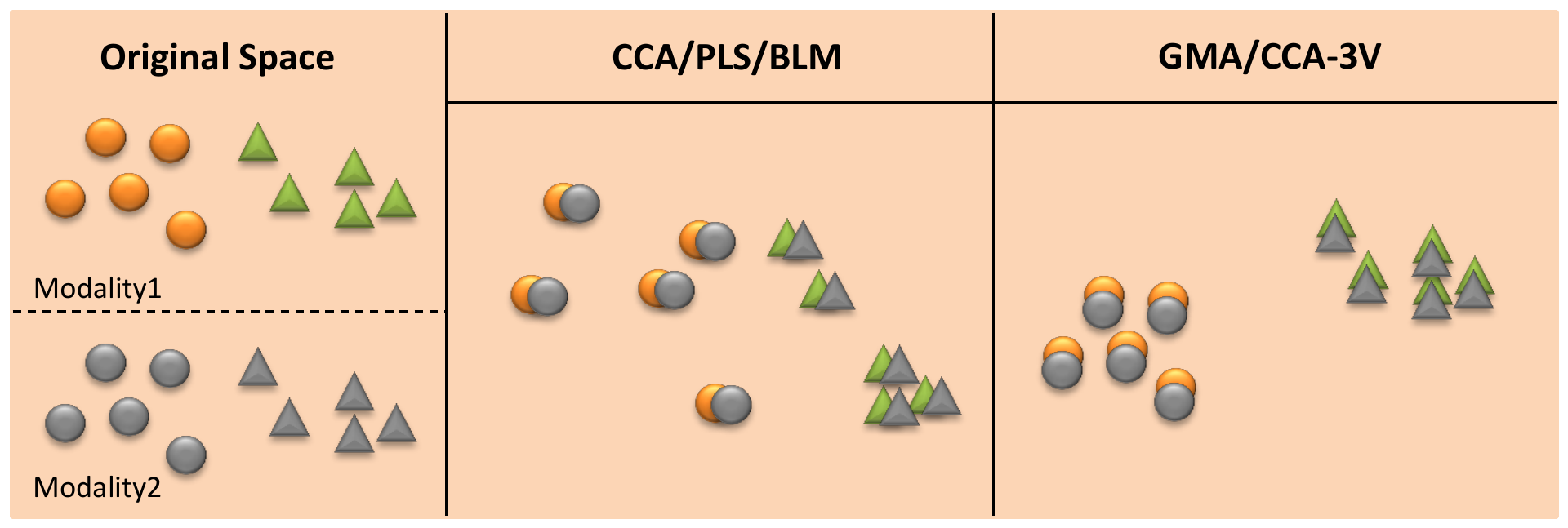}
    \caption{The difference between unsupervised subspace learning methods (CCA/PLS/BLM) and supervised subspace learning methods (GMA/CCA-3V).
              The same shape indicates the same class, and the same color indicates the same modality.}
    \label{fig:GMA}
\end{figure}

\subsubsection{Subspace learning methods}

Figure \ref{fig:GMA} shows the difference between unsupervised subspace learning methods and supervised subspace learning methods.
Supervised subspace learning methods enforce different-class samples to be mapped far apart while the same-class samples lie as close as possible.
To obtain more discriminative subspace, several works extend CCA to supervised subspace learning methods.

Sharma et al. \cite{ref32} present a supervised extension of CCA, called Generalized Multiview Analysis (GMA).
The optimal projection directions ${{\hat {\bf{v}}}_1}$, ${{\hat {\bf{v}}}_2}$ are obtained as
\begin{equation}\label{eq:GMA}
\begin{array}{l}
 [{{\hat {\bf{v}}}_1},{{\hat {\bf{v}}}_2}] = \mathop {\arg \max }\limits_{{{\bf{v}}_1},{{\bf{v}}_2}} {\bf{v}}_1^T{A_1}{{\bf{v}}_1} + \mu {\bf{v}}_2^T{A_2}{{\bf{v}}_2} + 2\alpha {\bf{v}}_1^T{Z_1}Z_2^T{{\bf{v}}_2} \\
 {\rm{     ~~~~~~~~~~~~~~~   }}s.t.{\rm{ ~~ }}{\bf{v}}_1^T{B_1}{{\bf{v}}_1} + \gamma {\bf{v}}_2^T{B_2}{{\bf{v}}_2} = 1 \\
 \end{array}
\end{equation}
It extends Linear Discriminant Analysis (LDA) and Marginal Fisher Analysis (MFA)
to their multiview counterparts, i.e., Generalized Multiview LDA (GMLDA) and Generalized Multiview MFA (GMMFA),
and apply them to deal with the cross-media retrieval problem. GMLDA and GMMFA take the semantic category into account, which has
obtained promising results.
Rasiwasia et al. \cite{rasiwasia2014cluster} present cluster canonical correlation analysis (cluster-CCA) for joint
dimensionality reduction of two modalities of data.
The cluster-CCA problem is formulated as

\begin{equation}\label{eq:cluster-cca}
\rho  = \mathop {\max }\limits_{{{\bf{w}}_x},{{\bf{w}}_y}} \frac{{{\bf{w}}_x^T{\Sigma_{xy}}{{\bf{w}}_y}}}{{\sqrt {{\bf{w}}_x^T{\Sigma_{xx}}{{\bf{w}}_x}} \sqrt {{\bf{w}}_y^T{\Sigma_{yy}}{{\bf{w}}_y}} }}
\end{equation}
where the covariance matrices ${{\Sigma_{xy}}}$, ${{\Sigma _{xx}}}$ and ${{\Sigma _{yy}}}$ are defined as

\begin{equation}\label{eq:cluster-cca2}
{\Sigma _{xy}} = \frac{1}{M}\sum\limits_{c = 1}^C {\sum\limits_{j = 1}^{|{{\bf{X}}_c}|} {\sum\limits_{k = 1}^{|{{\bf{Y}}_c}|} {{\bf{x}}_j^c{\bf{y}}_k^{c'}} } }
\end{equation}

\begin{equation}\label{eq:cluster-cca3}
{\Sigma _{xx}} = \frac{1}{M}\sum\limits_{c = 1}^C {\sum\limits_{j = 1}^{|{{\bf{X}}_c}|} {|{{\bf{Y}}_c}|{\bf{x}}_j^c{\bf{x}}_j^{c'}} }
\end{equation}

\begin{equation}\label{eq:cluster-cca4}
{\Sigma _{yy}} = \frac{1}{M}\sum\limits_{c = 1}^C {\sum\limits_{k = 1}^{|{{\bf{Y}}_c}|} {|{{\bf{X}}_c}|{\bf{y}}_k^c{\bf{y}}_k^{c'}} }
\end{equation}
where $M = \sum\nolimits_{c = 1}^C {|{{\bf{X}}_c}|} |{{\bf{Y}}_c}|$ is the total number of pairwise correspondences.
Cluster-CCA is able to learn discriminant low dimensional representations
that maximize the correlation between the two modalities of data while segregating the different classes on the learned space.
Gong et al. \cite{gong2014multi} propose a novel three-view CCA (CCA-3V) framework, which explicitly incorporates the dependence of
visual features and text on the underlying semantics. To better understand the CCA-3V, the objective function is given as below:
\begin{equation}\label{eq:cca-3v}
\begin{array}{l}
 \mathop {\min }\limits_{{{\bf{W}}_1},{{\bf{W}}_2},{{\bf{W}}_3}} \left\| {{\varphi _1}({\bf{V}}){{\bf{W}}_1} - {\varphi _2}({\bf{T}}){{\bf{W}}_2}} \right\|_F^2 +  \\
 \left\| {{\varphi _1}({\bf{V}}){{\bf{W}}_1} - {\varphi _3}({\bf{C}}){{\bf{W}}_3}} \right\|_F^2 + \left\| {{\varphi _2}({\bf{T}}){{\bf{W}}_2} - {\varphi _3}({\bf{C}}){{\bf{W}}_3}} \right\|_F^2 \\
 \end{array}
\end{equation}
where ${{\varphi _1}({\bf{V}})}$, ${{\varphi _1}({\bf{T}})}$, and ${{\varphi _1}({\bf{C}})}$ represent
embedding vectors from visual view, text view and semantic class view, respectively.
${\bf{W}}_1$, ${\bf{W}}_2$, and ${\bf{W}}_3$ are the learned projections for each view.
Furthermore, a distance function specially adapted to CCA that improves the accuracy of
retrieval in the embedded space is adopted.
Ranjan et al. \cite{Ranjan2015mlCCA} introduce multi-label Canonical Correlation Analysis (ml-CCA),
an extension of CCA, for learning shared subspaces by considering the high level semantic information
in the form of multi-label annotations.
They also present Fast ml-CCA, a computationally efficient version of ml-CCA,
which is able to handle large scale datasets.
Jing et al. \cite{jing2014intra} propose a novel multi-view feature learning approach based on
intra-view and inter-view supervised correlation analysis (I$^2$SCA). It explores
the useful correlation information of samples within each view and between all views.

Besides supervised CCA-based methods, Lin and Tang \cite{ref19} propose
a common discriminant feature extraction (CDFE) method to learn a common feature subspace where
the difference of within scatter matrix and between scatter matrix is maximized.
Mao et al. \cite{ref64} introduce a method for cross media retrieval, named parallel field alignment retrieval (PFAR), which integrates
a manifold alignment framework from the perspective of vector fields.
Zhai et al. \cite{zhai2014learning} propose a novel feature learning algorithm for cross-modal data, named Joint Representation
Learning (JRL). It can explore the correlation and semantic information in a unified optimization framework.

Wang et al. \cite{ref52} propose a novel regularization framework for the cross-modal matching problem, called LCFS (Learning Coupled Feature Spaces).
It unifies coupled linear regressions, $\ell_{21}$-norm
and trace norm into a generic minimization formulation so that subspace learning
and coupled feature selection can be performed simultaneously.
Furthermore, they extend this framework to more than two-modality case in \cite{Wang2016JFSSL},
 where the extension version is called JFSSL (Joint Feature Selection and Subspace Learning).
The main extensions are summarized as follows: 1) they propose a multimodal graph to better model
the similarity relationships among different modalities of data, which is demonstrated to outperform the low rank
constraint in terms of both computational cost and retrieval performance.
2) Accordingly, a new iterative algorithm is proposed to solve the modified
objective function and the proof of its convergence is given.

Inspired by the idea of (semi-)coupled dictionary learning,
Zhuang et al. \cite{zhuang2013supervised} bring coupled dictionary learning into supervised sparse coding for cross-modal retrieval,
which is called Supervised coupled dictionary learning with group structures for multi-modal retrieval (SliM$^2$).
It can utilize the class information to jointly learn discriminative multi-modal dictionaries
as well as mapping functions between different modalities.
The objective function is formulated as follows:
\begin{equation}\label{eq:SliM}
\begin{array}{l}
 \min \sum\limits_{m = 1}^M {\left\| {{{\bf{X}}^{(m)}} - {{\bf{D}}^{(m)}}{{\bf{A}}^{(m)}}} \right\|_F^2}  + \sum\limits_{m = 1}^M {\sum\limits_{l = 1}^J {{\lambda _m}{{\left\| {{\bf{A}}_{:,{\Omega _l}}^m} \right\|}_{1,2}}} }  \\
  + \beta \sum\limits_{m = 1}^M {\sum\limits_{n \ne m} {\left\| {{{\bf{A}}^{(n)}} - {{\bf{W}}^{(m)}}{{\bf{A}}^{(m)}}} \right\|_F^2} }  + \gamma \sum\limits_{m = 1}^M {\left\| {{{\bf{W}}^{(m)}}} \right\|_F^2}  \\
 s.t.{\rm{    }}\left\| {d_k^{(m)}} \right\| \le 1,{\rm{     }}\forall k,\forall m, \\
 \end{array}
\end{equation}
where ${{\bf{A}}_{:,{\Omega _l}}^m}$ is the coefficient matrix associated to those
intra-modality data belonging to the $l$-th class. As shown above,
data in the $m$-th modality space can be mapped into the $n$-th modality space by the learned ${{{\bf{W}}^{(m)}}}$
according to ${\left\| {{{\bf{A}}^{(n)}} - {{\bf{W}}^{(m)}}{{\bf{A}}^{(m)}}} \right\|_F^2}$,
therefore, the computation of cross-modal similarity is achieved.

\subsubsection{Topic models}

Based on Document Neural Autoregressive Distribution Estimator (DocNADE), Zhen et al. \cite{zheng2014topic} propose
a supervised extension of DocNADE, to learn a joint representation from image visual words, annotation words and class
label information.

Liao et al. \cite{liao2014nonparametric} present a nonparametric Bayesian
upstream supervised (NPBUS) multi-modal topic model for analyzing multi-modal data.
The NPBUS model allows flexible learning of correlation
structures of topics with individual modality and between
different modalities. The model becomes more discriminative
via incorporating upstream supervised information shared
by multi-modal data. Furthermore, it is capable to automatically
determine the number of latent topics in each modality.

Wang et al. \cite{wang2014multi} propose a supervised multi-modal mutual topic reinforce modeling (M$^3$R)
approach for cross-media retrieval, which seeks to build a joint cross-modal probabilistic graphical model for discovering the mutually
consistent semantic topics via appropriate interactions between
model factors (e.g., categories, latent topics and observed multi-modal data).

\subsubsection{Deep learning methods}

Wang et al. \cite{Wang15DeepMapping} propose a regularized deep neural network (RE-DNN) for
semantic mapping across modalities. They design and implement a 5-layer neural network for mapping visual and textual features into a common semantic
space such that the similarity between different modalities can be measured.

Li et al. \cite{Li15DeepSpace} propose a deep learning method to address the cross-media retrieval problem with multiple labels.
The proposed method is supervised, and the correlation between two modalities can
be built according to their shared ground truth probability vectors.
Both two networks have two hidden layers and one output layer, and the squared loss is employed as the loss function.

Wei et al. \cite{Wei16NewBaseline} propose a deep semantic matching method to
address the cross-modal retrieval problem for  samples
which are annotated with one or multiple labels.

Castrejon et al. \cite{Castrejon2016Aligned} present two approaches to regularize cross-modal
convolutional networks so that the intermediate representations
are aligned across modalities. The focus of this work is to learn cross-modal
representations when the modalities are significantly
different (e.g., text and natural images) and have category supervision.

Wang et al. \cite{Wang16EffectiveDeep} propose a supervised Multi-modal Deep Neural Network (MDNN) method, which consists of a deep convolutional neural network (DCNN) model and
a neural language model (NLM) to learn mapping functions
for the image modality and the text modality respectively.
It exploits the label information, and thus can learn robust mapping functions against noisy input data.

\section{Binary Representation Learning}

\begin{figure*}
    \centering
    \includegraphics[width=0.8\textwidth]{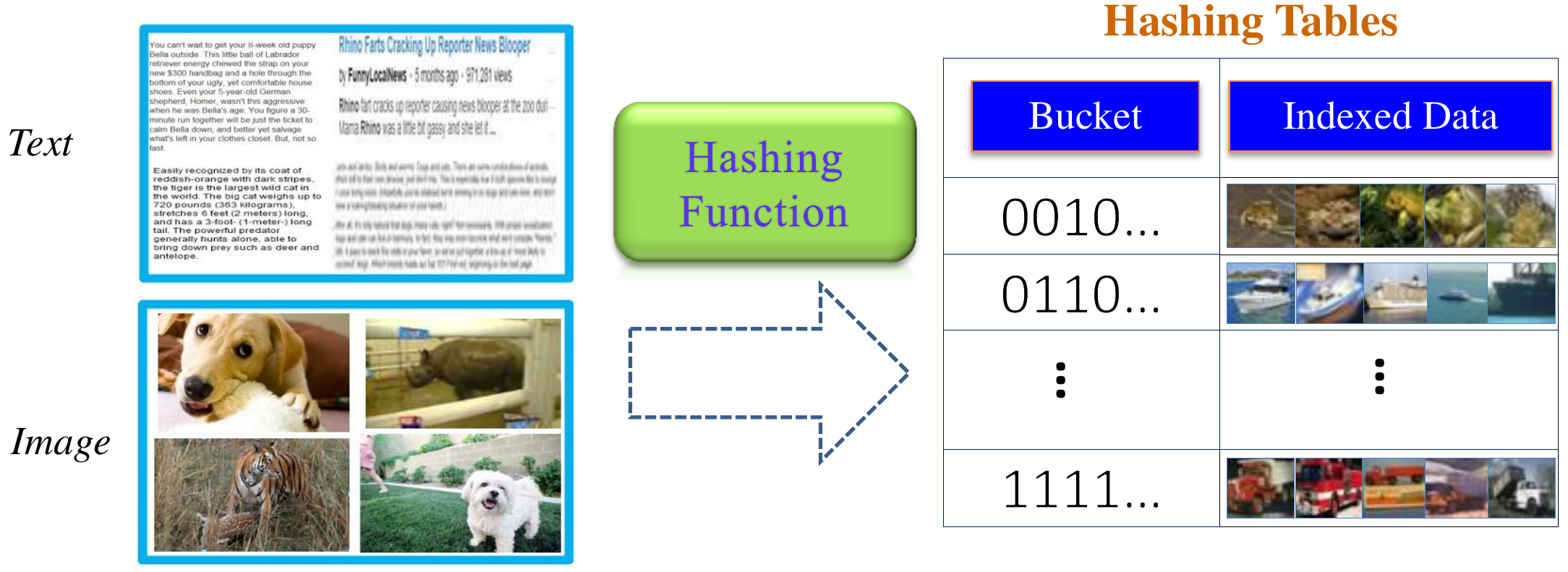}
    \caption{The basic idea of cross-modal hashing, which projects different modalities of data into a common Hamming
             space for fast retrieval.}
    \label{fig:hash}
\end{figure*}

Most existing real-valued cross-modal retrieval techniques are based on the brute-force linear search, which is time-consuming
for large scale data. A practical way to speed up the similarity search is binary representation learning, which is referred as hashing.
Existing hashing methods can be categorized into uni-modal hashing, multi-view hashing and cross-modal hashing.
Representative hashing methods on single modal data include spectral hashing (SH) \cite{Weiss2008SH}, self-taught hashing \cite{Zhang2010STH},
iterative quantization hashing (ITQ) \cite{Gong2011ITQ}, and so on.
The approaches mentioned above focus on learning hash functions for data objects with
homogeneous features. However, in real world applications, we often extract multiple types of features with different properties from data objects.
Accordingly, multi-view hashing methods (e.g., CHMIS \cite{Zhang2011CHMIS} and MFH \cite{Song2011MFH})
leverage information contained in different features to learn more accurate hash codes.

Cross-modal hashing methods aim to discover correlations among different modalities of data
to enable cross-modal similarity search. They project different modalities of data into a common Hamming
space for fast retrieval (as shown in Figure \ref{fig:hash}).
Similarly, cross-modal hashing methods can be categorized into:
1) unsupervised methods, 2) pairwise based methods, and 3) supervised methods.
To the best of our knowledge, there is little literature on rank based cross-modal hashing.
According to the learnt hash function being linear or nonlinear, the cross-modal hashing method
can be further divided into two categories: linear modeling and nonlinear modeling.
Linear modeling methods aim to learn linear functions to obtain hash codes.
Whereas, nonlinear modeling methods learn the hash codes in a nonlinear manners.

\subsection{Unsupervised methods}

\subsubsection{Linear modeling}

In \cite{ref14}, Kumar et al. propose a cross-view hashing (CVH),
which extends spectral hashing \cite{Weiss2008SH} from traditional uni-modal setting to
the multi-modal scenario. The hash functions map similar objects to similar codes across
views, and thus enable cross-view similarity search. The objective function is:
\begin{equation}\label{eq:CVH}
\begin{array}{l}
 {\rm{minimize}}:{\rm{ }}\bar d = \sum\limits_{i = 1}^n {\sum\limits_{j = 1}^n {{{\bf{W}}_{ij}}{d_{ij}}} }  \\
 {\rm{subject to}}:{\rm{ }}{{\rm{{\bf{Y}}}}^{(k)}}e = 0,for{\rm{ ~}}k = 1,...,K \\
 {\rm{   ~~~~~~~~~~~~~    }}\frac{1}{{Kn}}\sum\limits_{k = 1}^K {{{\bf{Y}}^{(k)}}{{\bf{Y}}^{(k)}}^T}  = {{\bf{I}}_d}, \\
 {\rm{    ~~~~~~~~~~~~~ }}{\bf{Y}}_{ij}^{(k)} \in \{  - 1,1\} ,for~{\rm{ }}k = 1,...,K \\
 \end{array}
\end{equation}
where $K$ is the total number of views, and $d_{ij}$ is the Hamming distance between objects $i$ and $j$ summed over all $K$ views:
\begin{equation}\label{eq:ddd}
{d_{ij}} = \sum\limits_{k = 1}^K {d({\bf{y}}_i^{(k)},{\bf{y}}_j^{(k)})}  + \sum\limits_{k = 1}^K {\sum\limits_{k' > k}^K {d({\bf{y}}_i^{(k)},{\bf{y}}_j^{(k')})} }
\end{equation}
Actually, CCA can be viewed as a special case of CVH by setting ${\bf{W}}={\bf{I}}$.

Rastegari et al. \cite{Rastegari2013PDH} propose a predictable dual-view hashing (PDH) algorithm for two-modalities.
They formulate an objective function to maintain the predictability of binary codes
and optimize the objective function by applying an iterative method based on block coordinate descent.

Ding et al. \cite{ding2014collective} propose a novel hashing method, which is referred to Collective
Matrix Factorization Hashing (CMFH). CMFH assumes that all modalities of an instance
generate identical hash codes. It learns unified hash codes by collective matrix factorization with
a latent factor model from different modalities of one instance. The objective function is:
\begin{equation}\label{eq:CMFH}
\begin{array}{l}
 \mathop {\min }\limits_{{{\bf{U}}_1},{{\bf{U}}_2},{{\bf{P}}_1},{{\bf{P}}_2},{\bf{V}}} \lambda \left\| {{{\bf{X}}^{(1)}} - {{\bf{U}}_1}{\bf{V}}} \right\|_F^2 + (1 - \lambda )\left\| {{{\bf{X}}^{(2)}} - {{\bf{U}}_2}{\bf{V}}} \right\|_F^2 \\
 {\rm{   ~~~~~~~~~~~~~  }} + \mu \left( {\left\| {{\bf{V}} - {{\bf{P}}_1}{{\bf{X}}^{(1)}}} \right\|_F^2 + \left\| {{\bf{V}} - {{\bf{P}}_2}{{\bf{X}}^{(2)}}} \right\|_F^2} \right) \\
 {\rm{     ~~~~~~~~~~~~~            }} + \gamma R({\bf{V}},{{\bf{P}}_1},{{\bf{P}}_2},{{\bf{U}}_1},{{\bf{U}}_2}) \\
 \end{array}
\end{equation}
where ${{{\bf{X}}^{(1)}}}$ and ${{{\bf{X}}^{(2)}}}$ represent two modalities of data, and
${\bf{V}}$ represents the latent semantic representations.
${{{\bf{P}}^{(1)}}}$ and ${{{\bf{P}}^{(2)}}}$ are the learned projections.

Zhou et al. \cite{zhou2014latent} propose a novel Latent Semantic Sparse Hashing (LSSH) to
perform cross-modal similarity search.
In particular, LSSH uses Sparse Coding to capture the salient structures of images:
\begin{equation}\label{eq:lssh1}
{O_{sc}}({\bf{B}},{\bf{S}}) = \left\| {{\bf{X}} - {\bf{BS}}} \right\|_F^2 + \sum\limits_{i = 1}^n {\lambda |{s_i}{|_1}}
\end{equation}
and uses Matrix Factorization to learn the latent concepts from text:
\begin{equation}\label{eq:lssh2}
{O_{mf}}({\bf{U}},{\bf{A}}) = \left\| {{\bf{Y}} - {\bf{UA}}} \right\|_F^2
\end{equation}
Then the learned latent semantic features are mapped to a joint abstraction space.
\begin{equation}\label{eq:lssh3}
{O_{cc}}({\bf{R}}) = \left\| {{\bf{A}} - {\bf{RS}}} \right\|_F^2
\end{equation}
The overall objective function is
\begin{equation}\label{eq:lssh4}
\begin{array}{l}
 \mathop {\min }\limits_{{\bf{B}},{\bf{A}},{\bf{R}},{\bf{U}},{\bf{S}}} O({\bf{B}},{\bf{A}},{\bf{R}},{\bf{U}},{\bf{S}}) = {O_{sc}} + {O_{mf}} + {O_{cc}} \\
 s.t.{\left\| {{{\bf{B}}_{.i}}} \right\|^2} \le 1,{\left\| {{{\bf{U}}_{.j}}} \right\|^2} \le 1,{\left\| {{{\bf{R}}_{.t}}} \right\|^2} \le 1,\forall i,j,t \\
 \end{array}
\end{equation}

Song et al. \cite{song2013inter} propose a novel inter-media hashing (IMH) model to transform multimodal data
into a common Hamming space. This method explores the inter-media consistency and intra-media consistency to
derive effective hash codes, based on which hash functions are learnt to efficiently
map new data points into the Hamming space. To learn the hash codes ${\bf{F}}^1$ and ${\bf{F}}^2$ for two modalities respectively,
the objective function is defined as follows:
\begin{equation}\label{eq:IMH}
\begin{array}{l}
 \mathop {\min }\limits_{{{\bf{F}}^1},{{\bf{F}}^2},{{\bf{W}}^1},{{\bf{W}}^2}} {\rm{ }}\lambda \sum\limits_{g = 1}^2 {\sum\limits_{p,q = 1}^{{N_g}} {{{({{\bf{A}}^g})}_{pq}}\left\| {f_p^g - f_q^g} \right\|_2^2} }  \\
 {\rm{              }} + Tr\left( {{{({{\bf{S}}^1}{{\bf{F}}^1} - {{\bf{S}}^2}{{\bf{F}}^2})}^T}{\bf{U}}({{\bf{S}}^1}{{\bf{F}}^1} - {{\bf{S}}^2}{{\bf{F}}^2})} \right) \\
 {\rm{               }} + \sum\limits_{g = 1}^2 {\left( {\left\| {{{({{\bf{X}}^g})}^T}{{\bf{W}}^g} - {{\bf{F}}^g}} \right\|_F^2 + \beta \left\| {{{\bf{W}}^g}} \right\|_F^2} \right)}  \\
 {\rm{   }}s.t.{\rm{       }}\left\{ \begin{array}{l}
 f_i^g \in {\{  - 1,1\} ^s} \\
 {({{\bf{F}}^1})^T}({{\bf{F}}^1}) = {I_s} \\
 \end{array} \right. \\
 \end{array}
\end{equation}
The first term models intra-modal consistency, the second term models inter-modal consistency,
and the third term learns the hash functions ${\bf{W}}^g$ to generate hash codes for new data.

Zhu et al. \cite{Zhu2013LCMH} propose a novel hashing method, named linear cross-modal hashing (LCMH),
to enable scalable indexing for multimedia search.
This method achieves a linear time complexity to the training data size in the training phase.
The key idea is to partition the training data of each modality into $k$ clusters,
and then represent each training data point with its distances to $k$ centroids of the clusters for preserving the intra-similarity in each modality.
To preserve the inter-similarity among data points across different modalities,
they transform the derived data representations into a common binary subspace.

\subsubsection{Nonlinear modeling}

The hash functions learned by most existing cross-modal hashing methods are linear.
To capture more complex structure of the multimodal data, nonlinear hash function learning is studied recently.
Based on stacked auto-encoders, Wang et al. \cite{wang2014effective} propose an effective nonlinear mapping mechanism
for multi-modal retrieval, called Multi-modal Stacked Auto-Encoders (MSAE).
Mapping functions are learned by optimizing a new objective function, which captures both intra-modal and inter-modal
semantic relationships of data from heterogeneous sources effectively.
The stacked structure of MSAE enables the method to learn nonlinear projections rather than linear projections.

Wang et al. \cite{Wang15DMHOR} propose a Deep Multimodal Hashing with Orthogonal
Regularization (DMHOR)
to learn accurate and compact multimodal representations. The method can better
capture the intra-modality and inter-modality correlations to
learn accurate representations. Meanwhile, in order to make the
representations compact and reduce redundant information lying in the codes,
an orthogonal regularizer is imposed on the learned weighting matrices.

\subsection{Pairwise based methods}

\subsubsection{Linear modeling}

To the best of our knowledge, Bronstein et al. \cite{ref13} propose the first cross-modal hashing method, called cross-modal
similarity sensitive hashing (CMSSH). CMSSH learns hash functions for the bimodal case in a standard boosting manner.
Specifically, given two modalities of data sets, CMSSH learns two groups of hash
functions to ensure that if two data points (with different modalities) are relevant, their corresponding hash codes
are similar and otherwise dissimilar.
However, CMSSH only preserves the inter-modality correlation but ignores the intra-modality similarity.

Zhen et al. \cite{zhen2012co} propose a novel multimodal hash function
learning method, called Co-Regularized Hashing (CRH), based on a boosted co-regularization framework.
The hash functions for each bit of hash codes are
learned by solving DC (difference of convex functions) programs.
To learn projections ${{{\bf{w}}_x}}$ and ${{{\bf{w}}_y}}$ from multimodal data, the objective function is defined as:
\begin{equation}\label{eq:crh}
O = \frac{1}{I}\sum\limits_{i = 1}^I {\ell _i^x}  + \frac{1}{J}\sum\limits_{j = 1}^J {\ell _j^y}  + \gamma \sum\limits_{n = 1}^N {{\omega _n}\ell _n^*}  + \frac{{{\lambda _x}}}{2}{\left\| {{{\bf{w}}_x}} \right\|^2} + \frac{{{\lambda _y}}}{2}{\left\| {{{\bf{w}}_y}} \right\|^2}
\end{equation}
where ${\ell _i^x}$ and ${\ell _j^y}$ are intra-modality loss terms for modalities $X$ and $Y$, defined as follows:
\begin{equation}\label{eq:crh1}
\ell _i^x = {\left[ {1 - |{\bf{w}}_x^T{{\bf{x}}_i}|} \right]_ + },{\rm{  ~~~~  }}\ell _j^y = {\left[ {1 - |{\bf{w}}_y^T{{\bf{y}}_j}|} \right]_ + }
\end{equation}
where ${[a]_ + }$ is equal to $a$ if $a \ge 0$ and $0$ otherwise.
The inter-modality loss term ${\ell _n^*}$ is defined as follows:
\begin{equation}\label{eq:crh2}
\ell _n^* = {s_n}d_n^2 + (1 - {s_n})\tau ({d_n})
\end{equation}
where ${d_n} = {\bf{w}}_x^T{{\bf{x}}_{{a_n}}} - {\bf{w}}_y^T{{\bf{y}}_{{b_n}}}$ and
$\tau (d)$ is the smoothly clipped inverted squared deviation function in Eq.\ref{eq:multi-npp3} \cite{ref38}.
The learning for multiple bits proceeds via a boosting procedure so that the bias introduced
by the hash functions can be sequentially minimized.

Hu et al. \cite{hu2014iterative} propose a  multi-view hashing algorithm for cross-modal retrieval, called
Iterative Multi-View Hashing (IMVH). IMVH aims to learn discriminative hashing functions for mapping multi-view
datum into a shared hamming space. It not only preserves the within-view similarity, but also incorporates
the between-view correlations into the encoding scheme, where it
maps the similar points to be close and push apart the dissimilar ones.

The cross-modal hashing methods usually assume that the hashed data reside in a common
Hamming space. However, this may be inappropriate, especially when the modalities are quite different.
To address this problem, Ou et al. \cite{ou2013comparing} propose a novel Relation-aware
Heterogeneous Hashing (RaHH), which provides a general
framework for generating hash codes of data entities from multiple heterogeneous domains.
Unlike some existing cross-modal hashing methods that map heterogeneous data into a
common Hamming space, the RaHH approach constructs a
Hamming space for each type of data entities, and learns optimal mappings ${{\bf{W}}^{pq}}$ between them simultaneously.
The RaHH framework encodes both homogeneous and heterogeneous relationships between the data entities to learn hash codes ${{\bf{H}}^p}$.
Specifically, the homogeneous loss is
\begin{equation}\label{eq:rahh1}
{J^{ho}}(\{ {{\bf{H}}^p}\} ) = \frac{1}{2}\sum\limits_{p = 1}^P {\sum\limits_{i,j = 1}^{{m_p}} {A_{ij}^p{{\left\| {{\bf{h}}_i^p - {\bf{h}}_j^p} \right\|}^2}} }
\end{equation}
where $A_{ij}^p$ indicates homogeneous relationship.
The heterogeneous loss is
\begin{equation}\label{eq:rahh2}
{J^{he}}(\{ {{\bf{H}}^p}\} ,\{ {{\bf{W}}^{pq}}\} ) = \sum\limits_{p \sim q} {\sum\limits_k {\sum\limits_{ij} {l_{ijk}^{pq}} } }  + \lambda {\left\| {{\bf{w}}_k^{pq}} \right\|^2}
\end{equation}
where ${p \sim q}$ indicates that modality $p$ has relationship with modality $q$, and the logistic loss is
\begin{equation}\label{eq:rahh3}
l_{ijk}^{pq} = \ln (1 + {e^{ - R_{ij}^{pq}H_{kj}^q{{({\bf{w}}_k^{pq})}^T}{\bf{h}}_i^p}})
\end{equation}
where ${R_{ij}^{pq}}$ indicates heterogeneous relationship.
To minimize the loss, ${H_{kj}^q}$ and ${{{({\bf{w}}_k^{pq})}^T}{\bf{h}}_i^p}$
need to be close for a large ${R_{ij}^{pq}}$.
Based on the similar idea, Wei et al. \cite{wei2014scalable} present a Heterogeneous Translated Hashing (HTH) method.
HTH simultaneously learns hash functions to embed heterogeneous media into different Hamming spaces
and translators to align these spaces.

Wu et al. \cite{Wu2015QCH} present a cross-modal hashing approach, called quantized correlation hashing (QCH),
which considers the quantization loss over domains and the relation between domains.
Unlike previous approaches that separate the optimization of the quantizer and the maximization of domain
correlation, this approach simultaneously optimizes both processes. The underlying relation between
the domains that describes the same objects is established via maximizing the correlation among
the hash codes across domains.

\subsubsection{Nonlinear modeling}

Zhen et al. \cite{Zhen2012MLBE} propose a probabilistic latent factor model, called
multimodal latent binary embedding (MLBE) to
learn hash functions for cross-modal retrieval. MLBE employs
a generative model to encode the intra-similarity and inter-similarity of data objects across multiple modalities. Based on
maximum a posteriori estimation, the binary latent factors are
efficiently obtained and then taken as hash codes in MLBE.
However, the optimization is easy to get trapped in local minima during learning, especially when the code length is large.

Zhai et al. \cite{zhai2013parametric} present a new parametric local multimodal
hashing (PLMH) method for cross-view similarity search. PLMH learns a set of hash functions to
locally adapt to the data structure of each modality.
Different local hash functions are learned at different locations of the input spaces,
therefore, the overall transformations of all points in each modality are locally linear but
globally nonlinear.

To learn nonlinear hash functions, Masci et al. \cite{masci2014multimodal} introduce
a novel learning framework for multimodal similarity-preserving hashing based on the
coupled siamese neural network architecture.
It utilizes similar pairs and dissimilar pairs for both intra- and inter-modality
similarity learning.
For modalities $X$ and $Y$, the objective function is defined as:
\begin{equation}\label{eq:deepHash}
\min {L_{XY}} + {L_X} + {L_Y}
\end{equation}
where two siamese networks are coupled by a cross-modal loss:
\begin{equation}\label{eq:deepHashXY}
\begin{array}{l}
 {L_{XY}} = \frac{1}{2}\sum\limits_{({\bf{x}},{\bf{y}}) \in {P_{XY}}} {\left\| {\xi ({\bf{x}}) - \eta ({\bf{y}})} \right\|_2^2}  \\
 {\rm{    ~~~~~ }} + \frac{1}{2}\sum\limits_{({\bf{x}},{\bf{y}}) \in {N_{XY}}} {\max {{\left\{ {0,{m_{XY}} - {{\left\| {\xi ({\bf{x}}) - \eta ({\bf{y}})} \right\|}_2}} \right\}}^2}}  \\
 \end{array}
\end{equation}
and the unimodal loss $L_X$ is
\begin{equation}\label{eq:deepHashX}
\begin{array}{l}
 {L_{XY}} = \frac{1}{2}\sum\limits_{({\bf{x}},{\bf{x'}}) \in {P_X}} {\left\| {\xi ({\bf{x}}) - \xi ({\bf{x'}})} \right\|_2^2}  \\
 {\rm{  ~~~~~   }} + \frac{1}{2}\sum\limits_{({\bf{x}},{\bf{x'}}) \in {N_X}} {\max {{\left\{ {0,{m_X} - {{\left\| {\xi ({\bf{x}}) - \xi ({\bf{x'}})} \right\|}_2}} \right\}}^2}}  \\
 \end{array}
\end{equation}
where $P$ and $N$ denote the similar pairs set and dissimilar pairs set, respectively.
$L_Y$ is similar to $L_X$.
The full multimodal version making use of inter- and intra-modal training data is called MM-NN in the original paper.
Unlike most existing cross-modality similarity learning approaches, the hashing functions are not limited to linear projections.
By increasing the number of layers in the network, mappings of the arbitrary complexity can be trained.

Cao et al. \cite{Cao16CHN} propose Correlation Hashing Network
(CHN), a new hybrid architecture for cross-modal hashing.
They jointly learn good image and text representations tailored
to hash coding and formally control the quantization error.


\subsection{Supervised methods}

\subsubsection{Linear modeling}

Zhang et al. \cite{zhang2014large} propose a multimodal hashing method, called semantic correlation
maximization (SCM), which integrates semantic labels into the hashing learning procedure.
This method uses label vectors to get semantic similarity matrix ${\bf{S}}$, and tries to reconstruct it through the learned hash codes.
Finally, the objective function is:
\begin{equation}\label{eq:SCM}
\mathop {\min }\limits_{f,g} \sum\limits_{i,j} {{{\left( {\frac{1}{c}f{{({{\bf{x}}_i})}^T}g({{\bf{y}}_j}) - {{\bf{S}}_{ij}}} \right)}^2}}
\end{equation}
To learn orthogonal projection, the objective function is reformulated as:
\begin{equation}\label{eq:SCM1}
\begin{array}{l}
 \mathop {\max }\limits_{{{\bf{W}}_x},{{\bf{W}}_y}} \left\| {({\bf{X}}{{\bf{W}}_x}){{({\bf{Y}}{{\bf{W}}_y})}^T} - c{\bf{S}}} \right\|_F^2 \\
 s.t.{\rm{ ~ }}{\bf{W}}_x^T{{\bf{X}}^T}{\bf{X}}{{\bf{W}}_x} = n{{\bf{I}}_c} \\
 {\rm{   ~~~~~    }}{\bf{W}}_y^T{{\bf{Y}}^T}{\bf{Y}}{{\bf{W}}_y} = n{{\bf{I}}_c}, \\
 \end{array}
\end{equation}
Furthermore, a sequential learning method (SCM-Seq) is proposed to learn the hash functions bit by bit
without imposing the orthogonality constraints.

Based on the dictionary learning framework,
Wu et al. \cite{wu2014sparse} develop an approach to obtain the sparse codesets for data objects across
different modalities via joint multi-modal dictionary learning,
which is called sparse multi-modal hashing (abbreviated as SM$^2$H).
In SM$^2$H, both intra-modality similarity and inter-modality
similarity are firstly modeled by a hypergraph, and then multi-modal
dictionaries are jointly learned by Hypergraph Laplacian sparse
coding. Based on the learned dictionaries, the sparse codeset
of each data object is acquired and conducted for multi-modal
approximate nearest neighbor retrieval using a sensitive Jaccard metric.
Similarly, Yu et al. \cite{yu2014discriminative} propose a discriminative coupled dictionary hashing (DCDH) method
to capture the underlying semantic information of the multi-modal data.
In DCDH, the coupled dictionary for each modality is learned
with the aid of class information. As a result, the coupled dictionaries not only preserve the intra-similarity and
inter-correlation among multi-modal data, but also contain
dictionary atoms that are semantically discriminative (i.e., data from the same category are reconstructed by the
similar dictionary atoms). To perform fast cross-media retrieval, hash functions are learned to map data from the
dictionary space to a low-dimensional Hamming space.

\subsubsection{Nonlinear modeling}

To capture more complex data structure,
Lin et al. \cite{Lin15SePH} propose a two-step supervised hashing method
termed SePH (Semantics-Preserving Hashing) for cross-view retrieval.
For training, SePH firstly transforms the given semantic affinities of training
data into a probability distribution and approximates it with
to-be-learnt hash codes in Hamming space via minimizing
the KL-divergence. Then in each view, SePH utilizes kernel
logistic regression with a sampling strategy to learn the nonlinear
projections from features to hash codes.
And for any unseen instance, predicted hash
codes and their corresponding output probabilities from observed
views are utilized to determine its unified hash code,
using a novel probabilistic approach.

Cao et al. \cite{Cao16CAH} propose a novel supervised cross-modal
hashing method, Correlation Autoencoder Hashing (CAH), to learn
discriminative and compact binary codes based on deep autoencoders.
Specifically, CAH jointly maximizes the feature correlation
revealed by bimodal data and the semantic correlation conveyed in
similarity labels, while embeds them into hash codes by nonlinear
deep autoencoders.

Jiang et al. \cite{JiangL16DeepCrossHash} propose a novel cross-modal hashing method,
called deep cross-modal hashing (DCMH), by integrating feature
learning and hash-code learning into the same framework. DCMH is an end-to-end learning
framework with deep neural networks, one for each modality, to perform feature learning from scratch.

\section{Multimodal datasets}

\begin{table*}
\begin{center}
\begin{tabular}{|l|c|c|c|c|c|c|c|c|c|}
\hline
Dataset       & Modality     & Number of Samples       & Image Features   & Text Features  & Number of Categories   \\
\hline\hline
 Wiki                & image/text     & 2,866          &  SIFT+BOW   &   LDA   &    10 \\
 INRIA-Websearch      & image/text     & 71,478       &   ---       &   ---     &   353 \\
 Flickr30K        & image/sentences      & 31,783         & ---    &   ---   &     ---  \\
 NUS-WIDE       & image/tags             & 186,577         &  6 types   &  tag occurrence feature    &   81    \\
 Pascal-VOC      & image/tags            &  9,963       &  3 types   &  tag occurrence feature    &  20   \\
\hline
\end{tabular}
\end{center}
\caption{The summarization of the multimodal datasets.}
\label{tab:Datasets}
\end{table*}

With the popularity of multimodal data, cross-modal
retrieval becomes an urgent and challenging problem.
To evaluate the performance of cross-modal retrieval algorithms,
researchers collect multimodal data to build up multimodal datasets.
Here, we introduce five commonly used datasets, i.e., Wikipedia, INRIA-Websearch,
Flickr30K, Pascal VOC, and NUS-WIDE datasets.

\begin{figure*}
    \centering
    \includegraphics[width=0.85\textwidth]{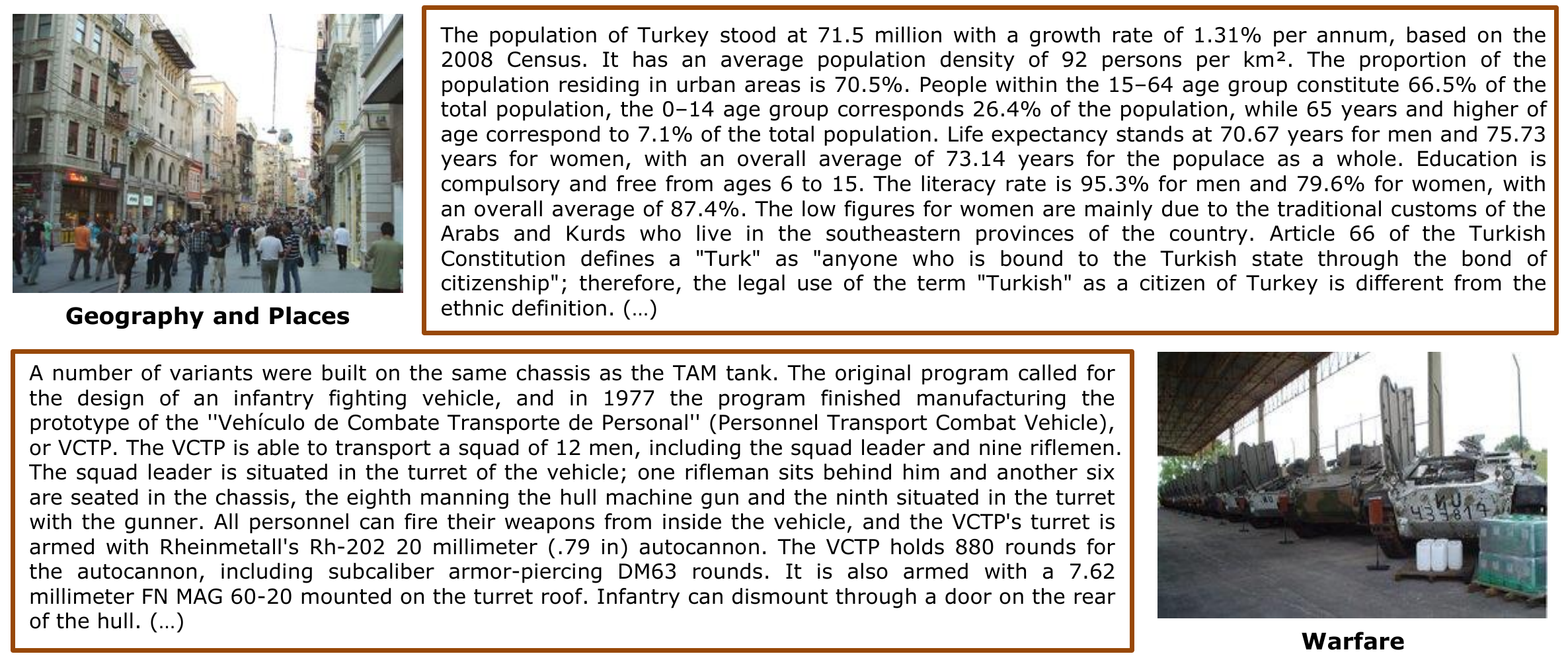}
    \caption{Two examples on the Wiki dataset. The text is an article describing the content of an image.}
    \label{fig:WikiSamples}
\end{figure*}

The \textit{Wiki image-text} dataset\footnote{http://www.svcl.ucsd.edu/projects/crossmodal/} \cite{ref06}:
it is generated from Wikipedia's ``featured article'',
which consists of 2866 image-text pairs. In each pair,
the text is an article describing people, places or some events and the image is closely related to the content of the article (as shown in Figure \ref{fig:WikiSamples}).
Besides, each pair is labeled with one of 10 semantic classes.
The representation of the text with 10 dimensions is derived from a latent Dirichlet allocation model \cite{ref42}.
The images are represented by the 128 dimensional SIFT descriptor histograms \cite{ref27}.

\begin{figure*}
    \centering
    \includegraphics[width=0.8\textwidth]{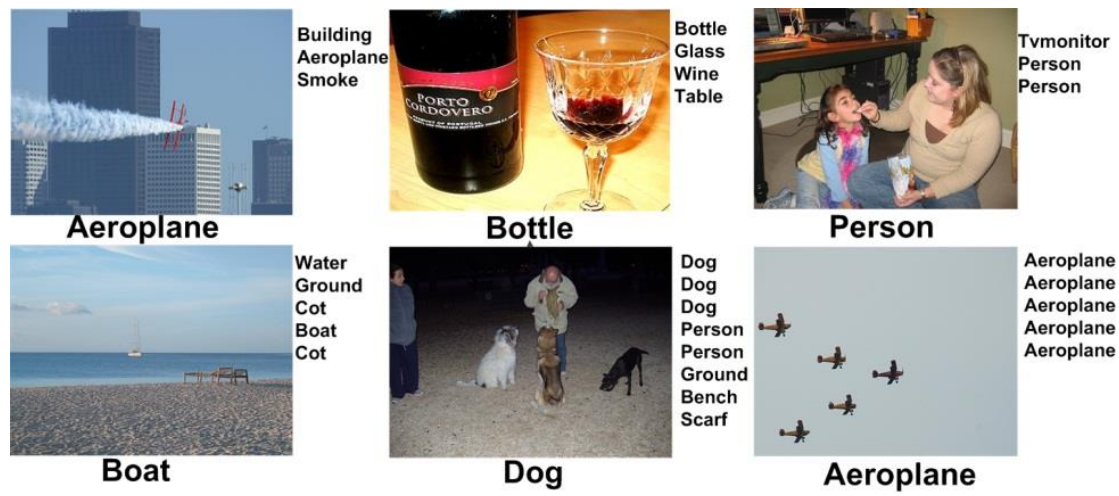}
    \caption{Examples on the Pascal VOC dataset. The tags describing the content of an image are regarded as the text modality.}
    \label{fig:VOCSamples}
\end{figure*}

The \textit{INRIA-Websearch} dataset\footnote{http://lear.inrialpes.fr/pubs/2010/KAVJ10/} \cite{Krapac2010INRIAWebsearchDataset}:
it contains 71,478 image-text pairs, which can be categorized into 353 different concepts
that include famous landmarks, actors, films, logos, etc.
Each concept comes with a number of images retrieved via Internet search,
and each image is marked as either relevant or irrelevant to its query concept.
The text modality consists of text surrounding images on web pages.
This dataset is very challenging because it contains a large number of classes.

The \textit{Flickr30K} dataset\footnote{http://shannon.cs.illinois.edu/DenotationGraph/} \cite{Young2014Flickr30}:
it is an extension of Flicker8K \cite{Hodosh2014Flickr8K},
which contains 31,783 images collected from different Flickr groups and
focuses on events involving people and animals.
Each image is associated with five sentences
independently written by native English speakers from Mechanical Turk.

The \textit{NUS-WIDE} dataset\footnote{http://lms.comp.nus.edu.sg/research/NUS-WIDE.htm} \cite{ref54}:
this dataset contains 186,577 labeled images.
Each image is associated with user tags, which can be taken as an image-text pair.
To guarantee that each class has abundant training samples, researchers generally select those pairs that belong to one of
the K (K=10, or 21) largest classes with each pair exclusively belonging to one of the selected classes.
Six types of low-level features are extracted from these images, including 64-D color histogram, 144-D color
correlogram, 73-D edge direction histogram, 128-D wavelet texture, 225-D block-wise color moments extracted over 5¡Á5
fixed grid partitions, and 500-D bag of words based on SIFT descriptions.
The textual tags are represented with 1000-dimensional tag occurrence feature vectors.

The \textit{Pascal VOC} dataset\footnote{http://www.cs.utexas.edu/~grauman/research/datasets.html} \cite{ref33}: it
consists of 5011/4952 (training/testing) image-tag pairs, which can be categorized into 20 different classes.
Some examples are shown in Figure \ref{fig:VOCSamples}.
Since some images are multi-labeled, researchers usually select images with only one object as the way in \cite{ref32}, resulting in 2808 training and 2841 testing data.
The image features include histograms of bag-of-visual-words, GIST and color \cite{ref33}, and the text features are 399-dimensional tag occurrence features.

These datasets generally contain two modalities of data, i.e., image and text.
Among those datasets, only the Wiki dataset is designed for cross-modal retrieval,
but the number of samples and categories it contains are small.
Another commonly used dataset is the NUS-WIDE dataset, especially in cross-modal hashing, due to its relative massiveness.
But the text modality in the NUS-WIDE dataset is user tag, which is simple and limited for text description.
The Flickr30K dataset is usually used for the image-sentence retrieval, which does not have category information.
So it is desirable to design a more general, large-scale multimodal dataset for future research,
which contains more modalities of data and more categories.

\section{Experiments}

In this section, we test the performance of different kinds of cross-modal retrieval methods.
Firstly, we introduce evaluation metrics,
then choose representative cross-modal retrieval methods for evaluation.
For the compared methods, we cite the reported results in the relevant literature.

\subsection{Evaluation metric}

To evaluate the performance of the cross-modal retrieval methods, two cross-modal retrieval tasks are conducted:
(1) Image query vs. Text database, (2) Text query vs. Image database.
More specifically, in testing phase, we take one modality of data of the testing set as the query set to retrieve another modality of data.
The cosine distance is adopted to measure the similarity of features.
Given an image (or text) query, the goal of each cross-modal task is to find the nearest neighbors from the text (or image) database.

The \textit{mean average precision (MAP)} \cite{ref06} is used to evaluate the overall performance of the tested algorithms.
To compute MAP, we first evaluate the average precision (AP) of a set of $R$ retrieved documents by
$
{\rm{AP}} = \frac{1}{T}\sum\nolimits_{r = 1}^R {P(r)\delta (r)}
$,
where $T$ is the number of relevant documents in the retrieved set, ${P(r)}$ denotes the precision of the top $r$ retrieved documents, and ${\delta (r)}=1$ if the $r$th
retrieved document is relevant (where 'relevant' means belonging to the class of the query)
and ${\delta (r)}=0$ otherwise. The MAP is then computed by averaging the AP values over all queries in the query set.
The larger the MAP, the better the performance.

For image-sentence retrieval, the commonly used metric is another one, which will be described in Section \ref{sec:flickr30k}.

\subsection{Comparison of real-valued representation learning methods}

For image-text retrieval, the \textit{Wiki image-text dataset} and the \textit{NUS-WIDE dataset}
are commonly used to evaluate performance.
For the real-valued representation learning methods, we choose three popular unsupervised methods (i.e., PLS \cite{ref18}, BLM \cite{ref32,ref30} and CCA \cite{ref06}),
two rank based methods (i.e., LSCMR \cite{ref55}, and Bi-LSCMR \cite{ref56})
and eight popular supervised methods (i.e., CDFE \cite{ref19}, GMLDA \cite{ref32}, GMMFA \cite{ref32},
CCA-3V \cite{gong2014multi}, SliM$^2$ \cite{zhuang2013supervised}, M$^3$R \cite{wang2014multi}, LCFS \cite{ref52} and JFSSL \cite{Wang2016JFSSL}).

For image-sentence retrieval, the \textit{Flickr30K} is the commonly used dataset for evaluation.
Deep learning methods are generally used for modeling the images and sentences.
For the deep learning methods, we choose four recently proposed methods:
DeViSE \cite{frome2013devise}, SDT-RNN \cite{socher2014grounded}, Deep Fragment \cite{Karpathy14DeepFragment} and End-to-end DCCA \cite{Yan2015DCCA}

\subsubsection{Results on Wiki and NUS-WIDE}

Tables \ref{tab:mapWiki} and \ref{tab:mapNUS} show the MAP scores
\footnote{For the Wiki and NUS-WIDE datasets, we use the features provided by the authors.
For PLS, BLM, CCA, CDFE, GMMFA, GMLDA, CCA-3V, LCFS and JFSSL,
we adopt the codes provided by the authors and use the setting in \cite{Wang2016JFSSL}.
For LSCMR, Bi-LSCMR, SliM$^2$, and M$^3$R, we cite the publicly reported results.}
achieved by
PLS \cite{ref18}, BLM \cite{ref32,ref30}, CCA \cite{ref06}, LSCMR \cite{ref55}, Bi-LSCMR \cite{ref56},
CDFE \cite{ref19}, GMMFA \cite{ref32}, GMLDA \cite{ref32},
CCA-3V \cite{gong2014multi}, SliM$^2$ \cite{zhuang2013supervised}, M$^3$R \cite{wang2014multi}, LCFS \cite{ref52} and JFSSL \cite{Wang2016JFSSL}.
For the experiments on the NUS-WIDE dataset, Principal Component Analysis (PCA) is first
performed on the original features to remove redundancy for methods PLS, BLM, CCA, CDFE, GMMFA, GMLDA, and CCA-3V.

We observe that the supervised learning methods (CDFE, GMMFA, GMLDA, CCA-3V, SliM$^2$, and M$^3$R, LCFS and JFSSL)
perform better than the unsupervised learning methods (PLS, BLM and CCA).
The reason is that PLS, BLM and CCA only care about pair-wise closeness in the common subspace,
but CDFE, GMMFA, GMLDA, CCA-3V, SliM$^2$, and M$^3$R, LCFS and JFSSL utilize class information to
obtain much better separation between classes in the common representation space.
Hence, it is helpful for cross-modal retrieval to learn a discriminative common representation space.
The rank based methods (LSCMR and Bi-LSCMR) achieve comparable results on the Wiki dataset, but perform worse on the
NUS-WIDE dataset. One of the reasons is that they only use a small set of data to generate rank lists on the NUS-WIDE dataset,
and the number of data used for training is not enough. Another reason is that the features (the bag of words
(BoW) representation with the TF-IDF weighting scheme for text, the bag-of-visual words (BoVW) feature for images) they used are not powerful enough.
LCFS and JFSSL performs the best, and one reason is that they perform feature selection on different feature spaces simultaneously.

\textbf{Results on the Wiki with different types of features:}
To evaluate the effect of different types of features,
we test the cross-modal retrieval performance with different types of features for images and texts on the Wiki dataset.
Here we cite the results in \cite{Wang2016JFSSL}.
Besides the features provided by the Wiki dataset itself, 4096-dimensional CNN(Convolutional Neural Networks) features for images are extracted by Caffe \cite{jia2014caffe}, and
5000-dimensional feature vectors for texts are extracted by using the bag of words representation with the TF-IDF weighting scheme.
Table \ref{tab:mapWikiFeats} shows the MAP scores of GMLDA, CCA-3V, LCFS and JFSSL with different types of features on the Wiki dataset.
PCA is performed on CNN and TF-IDF features for GMLDA and CCA-3V.
It can be seen that all methods achieve better results when using the CNN features.
This is because CNN features are more powerful for image representation, which has been proved in many fields.
Overall, better representation of data leads to better performance.

\begin{table}
\begin{center}
\begin{tabular}{|l|c|c|c|c|c|c|c|c|c|}
\hline
Methods       & Image query     & Text query       & Average   \\
\hline\hline
 PLS        & 0.2402     & 0.1633         & 0.2032      \\
 BLM        & 0.2562     & 0.2023         & 0.2293      \\
 CCA        & 0.2549     & 0.1846         & 0.2198      \\
 LSCMR        & 0.2021     & 0.2229         & 0.2125      \\
 Bi-LSCMR     & 0.2123     & 0.2528         & 0.2326      \\
 CDFE       & 0.2655    & 0.2059         & 0.2357      \\
 GMMFA       & 0.2750    & 0.2139         & 0.2445      \\
 GMLDA       & 0.2751    & 0.2098         &  0.2425    \\
 CCA-3V       & 0.2752    & 0.2242         &  0.2497    \\
 SliM$^2$     & 0.2548    & 0.2021         &  0.2285    \\
 M$^3$R        & 0.2298    & 0.2677         &  0.2488    \\
 LCFS       & 0.2798     & 0.2141        & 0.2470      \\
 JFSSL       & 0.3063     & 0.2275         & 0.2669      \\
\hline
\end{tabular}
\end{center}
\caption{MAP Comparison of different shallow real-valued representation learning methods on the Wiki dataset.}
\label{tab:mapWiki}
\end{table}

\begin{table}
\begin{center}
\begin{tabular}{|c|c|c|c|c|c|c|}
\hline
\multirow{2}{*}{Query}      &\multirow{2}{*}{Methods}    &\multicolumn{4}{c|}{Features (Image/Text)}\\
\cline{3-6}
   &     & {\tabincell{c}{SIFT/ \\ LDA}}  & {\tabincell{c}{CNN/ \\ LDA}}  & {\tabincell{c}{SIFT/ \\ TF-IDF}}  &{\tabincell{c}{CNN/ \\TF-IDF}}\\
\hline\hline
\multirow{4}{*}{Image}
                        &GMLDA       &   0.2751                 & 0.4084            & 0.2782     & 0.4455             \\ \cline{2-6}
                       &CCA-3V        &  0.2752                  &   0.4049         & 0.2862         & 0.4370             \\   \cline{2-6}
                        &LCFS        &  0.2798                 & 0.4132              &0.2978        & 0.4553             \\     \cline{2-6}
                        &JFSSL      &    0.3063                & 0.4279             & 0.3080          & 0.4670     \\  \cline{2-6}
\hline\hline
\multirow{4}{*}{Text}
                        &GMLDA      &    0.2098                & 0.3693             &   0.1925         &  0.3661        \\  \cline{2-6}
                        &CCA-3V     &   0.2242                & 0.3651                &   0.2238        &   0.3832                       \\  \cline{2-6}
                         &LCFS      &   0.2141                & 0.3845               &  0.2134             &  0.3978                   \\  \cline{2-6}
                         &JFSSL     &   0.2275               & 0.3957                 & 0.2257            &  0.4102         \\  \cline{2-6}
\hline
\end{tabular}
\end{center}
\caption{MAP comparison with different features on the Wiki dataset.}
\label{tab:mapWikiFeats}
\end{table}


\textbf{Results on the Wiki in three-modality case:}
To evaluate the performance of cross-modal retrieval methods in a three-modality case, we test several methods on the Wiki dataset.
Here we cite the results in \cite{Wang2016JFSSL}.
To the best of our knowledge, there are no three or more modalities of datasets available publicly in the recent literature.
The Wiki dataset contains two modalities of data: text and image.
We adopt the settings in \cite{Wang2016JFSSL}.
To simulate a three-modality setting, 4096-dimensional CNN (Convolutional Neural Networks) features of images are extracted
by Caffe \cite{jia2014caffe} as another virtual modality.
Here 128-dim SIFT histogram, 10-dim LDA feature and 4096-dimensional CNN features
are used as Modality A, Modality B and Modality C, respectively.
Table \ref{tab:3V-Wiki} shows the MAP comparison on the Wiki dataset in the three-modality case.
We can see that JFSSL outperforms the other methods in three cross-modal retrieval tasks.
This is mainly due to the fact that JFSSL is designed for the $N$-modality case, which can model the
correlations between different modalities more accurately in the three-modality case.
However, the other methods are designed for only the two-modality case,
and they are not suitable for the three-modality case.

From the above experiments, we draw the following conclusions:
\begin{itemize}
\item Supervised learning methods generally achieve better results than unsupervised learning methods.
The reason is that unsupervised methods only care about pair-wise closeness in the common subspace,
but supervised learning methods utilize class information to
obtain much better separation between classes in the common representation space.
Hence, it is helpful for cross-modal retrieval to learn a discriminative common representation.
\item For cross-modal retrieval, better features generally lead to better performance.
So it is beneficial for cross-modal retrieval to learn powerful representations for various modalities of data.
\item For algorithms designed for the two-modality case, it cannot be directly extended for more than two-modality case well.
So in more than two-modality case, they generally perform worse than the algorithms designed for more than two-modality methods.
\end{itemize}

\begin{table}
\begin{center}
\begin{tabular}{|l|c|c|c|c|c|c|c|c|c|}
\hline
Methods       & Image query     & Text query       & Average   \\
\hline\hline
 PCA+PLS        & 0.2752      & 0.2661         & 0.2706      \\
 PCA+BLM          & 0.2976     & 0.2809         & 0.2892      \\
 PCA+CCA        & 0.2872     & 0.2840         & 0.2856      \\
 LSCMR        & 0.1424     & 0.2491         & 0.1958      \\
 Bi-LSCMR     & 0.1453     & 0.2380         & 0.1917      \\
 PCA+CDFE      & 0.2595    & 0.2869         & 0.2732       \\
 PCA+GMMFA        & 0.2983    & 0.2939         & 0.2961      \\
 PCA+GMLDA      & 0.3243    & 0.3076         &  0.3159    \\
 PCA+CCA-3V        & 0.3513    & 0.3260         &  0.3386     \\
 SliM$^2$     & 0.3154    & 0.2924         &  0.3039    \\
 M$^3$R        & 0.2445    & 0.3044         &  0.2742    \\
 LCFS       & 0.3830     & 0.3460        & 0.3645     \\
 JFSSL       & 0.4035     & 0.3747         & 0.3891      \\
\hline
\end{tabular}
\end{center}
\caption{MAP Comparison of different shallow real-valued representation learning methods on the NUS-WIDE dataset.}
\label{tab:mapNUS}
\end{table}

\begin{table}
\begin{center}
\begin{tabular}{|l|c|c|c|c|c|c|c|c|c|}
\hline
Query       & Modality A     & Modality B       & Modality C   \\
\hline\hline
 PLS        & 0.1629     & 0.1653         & 0.2412      \\
 BLM        & 0.1673     & 0.2167         & 0.2607      \\
 CCA        & 0.1733     & 0.1722         & 0.2434      \\
 CDFE       & 0.1882    & 0.1836         & 0.2548      \\
 GMMFA       & 0.2005    & 0.1961         & 0.2551      \\
 GMLDA       & 0.1841    & 0.1700        &  0.2525    \\
 CCA-3V       & 0.2301    & 0.1720         &  0.2665    \\
 LCFS       & 0.2292     & 0.3065         & 0.3072     \\
 JFSSL       & 0.2636     & 0.3203         & 0.3354      \\
\hline
\end{tabular}
\end{center}
\caption{MAP Comparison of different shallow real-valued representation learning methods on the Wiki dataset in the three-modality case.}
\label{tab:3V-Wiki}
\end{table}

\subsubsection{Results on Flickr30K}
\label{sec:flickr30k}

For the Flickr30K dataset, We adopt the evaluation metrics in \cite{Karpathy14DeepFragment} for a fair comparison.
More specifically, for the image-sentece retrieval, we report the median rank (Med r)
of the closest ground truth result in the list,
as well as the R@K (with K = 1, 5, 10) that computes the fraction of times
the correct result being found among the top K items.
In contrast to R@K, a lower median rank indicates
a better performance.

Table \ref{tab:rankFlickr30K} shows the publicly reported results
achieved by DeViSE \cite{frome2013devise}, SDT-RNN \cite{socher2014grounded},
Deep Fragment \cite{Karpathy14DeepFragment} and End-to-end DCCA \cite{Yan2015DCCA} on the Flickr30K dataset.
End-to-end DCCA and Deep Fragment perform better than other methods.
The reason is that Deep Fragment breaks an image into objects and a sentence
into dependency tree relations, and maximises the explicit
alignment between the image fragments and text fragments.
For End-to-end DCCA, the used TF-IDF based text features and CNN based
visual features capture global properties of the two modalities
respectively. The alignment of the fragments in image
and text is implicitly considered by the CCA correlation
objective.

\begin{table*}
\begin{center}
\begin{tabular}{|c|c|c|c|c|c|c|c|c|c|}
\hline
\multirow{2}{*}{Methods}    &\multicolumn{4}{c|}{Sentence retrieval}    &\multicolumn{4}{c|}{Image retrieval}\\
\cline{2-9}
& {\tabincell{c}{R@1}}  & {\tabincell{c}{R@5}}  & {\tabincell{c}{R@10}}  &{\tabincell{c}{Med r}}   & {\tabincell{c}{R@1}}  & {\tabincell{c}{R@5}}  & {\tabincell{c}{R@10}}  &{\tabincell{c}{Med r}}\\
\hline\hline
DeViSE        &  4.5                 &   18.1        & 29.2         & 26     &   6.7                 & 21.9            & 32.7       & 25          \\   \cline{1-9}
SDT-RNN        &  9.6                 & 29.8             & 41.1        & 16      &   8.9                 & 29.8            & 41.1     & 16       \\     \cline{1-9}
Deep Fragment      & 14.2               & 37.7            & 51.3          & 10     &   10.2                 & 30.8           & 44.2     & 14    \\  \cline{1-9}
End-to-end DCCA       &   16.7                 & 39.3            & 52.9     & 8           &   12.6                 & 31.0            & 43.0     & 15    \\ \cline{1-9}
\hline
\end{tabular}
\end{center}
\caption{MAP comparison of different deep real-valued representation learning methods on the Flickr30K dataset.}
\label{tab:rankFlickr30K}
\end{table*}

\begin{table}
\begin{center}
\begin{tabular}{|c|c|c|c|c|c|c|}
\hline
\multirow{2}{*}{Tasks}      &\multirow{2}{*}{Methods}    &\multicolumn{4}{c|}{Code Length}\\
\cline{3-6}
                              &        & $K$=16           & $K$=32      & $K$=64     & $K$=128\\
\hline\hline
\multirow{8}{*}{\tabincell{c}{Image query \\ vs. \\ Text database}}
                        &CVH       &   0.1257                     & 0.1212             & 0.1215       & 0.1171         \\ \cline{2-6}
                         &IMH       &    0.1573                   & 0.1575            & 0.1568       &  0.1651         \\   \cline{2-6}
                       &LSSH        &   0.2141                  & 0.2216            & 0.2218       &  0.2211         \\   \cline{2-6}
                       &CMFH       &    0.2132                  & 0.2259            & 0.2362       &   0.2419        \\   \cline{2-6}
                        &CMSSH        &   0.1877                    & 0.1771         & 0.1646          & 0.1552         \\     \cline{2-6}
                        &SCM-Seq      &     0.2210                 & 0.2337          & 0.2442          &  0.2596      \\  \cline{2-6}
                         &SePH$_{km}$      &     0.2787                 & 0.2956           & 0.3064          & 0.3134       \\  \cline{2-6}
                         &MM-NN      &     --                 & 0.5750           & --          & --       \\  \cline{2-6}
\hline\hline
\multirow{8}{*}{\tabincell{c}{Text query \\ vs. \\ Image database}}
                        &CVH       &   0.1185                    & 0.1034            & 0.1024       &  0.0990        \\ \cline{2-6}
                        &IMH       &    0.1463                   & 0.1311            & 0.1290      &   0.1301        \\   \cline{2-6}
                       &LSSH        &    0.5031                   & 0.5224            & 0.5293       & 0.5346          \\   \cline{2-6}
                       &CMFH       &    0.4884                  & 0.5132             & 0.5269       &  0.5375         \\   \cline{2-6}
                        &CMSSH        &   0.1630                    & 0.1617         & 0.1539          &  0.1517        \\     \cline{2-6}
                        &SCM-Seq      &     0.2134                 & 0.2366           & 0.2479          &  0.2573      \\  \cline{2-6}
                         &SePH$_{km}$      &     0.6318                 & 0.6577          & 0.6646          &  0.6709      \\  \cline{2-6}
                         &MM-NN      &     --                & 0.2740          & --         &  --    \\  \cline{2-6}
\hline
\end{tabular}
\end{center}
\caption{MAP comparison of different cross-modal hashing methods on the Wiki dataset.}
\label{tab:mapWikiHash}
\end{table}

\subsection{Comparison of binary representation learning methods}

For cross-modal hashing methods, the \textit{Wiki image-text dataset} and the \textit{NUS-WIDE dataset}
are most commonly used datasets. We use the settings in \cite{Lin15SePH} and utilize the Mean Average Precision (MAP) as the evaluation metric.


Tables \ref{tab:mapWikiHash} and \ref{tab:mapNUSHash} show the publicly reported MAP scores
achieved by
unsupervised cross-modal hashing methods (CVH \cite{ref14}, IMH \cite{song2013inter}, LSSH \cite{zhou2014latent}, and CMFH \cite{ding2014collective}),
pairwise base methods (CMSSH \cite{ref13} and MM-NN \cite{masci2014multimodal}), supervised methods (SCM-Seq \cite{zhang2014large} and SePH$_{km}$ \cite{Lin15SePH})
on the Wiki and NUS-WIDE datasets, respectively.
MM-NN \footnote{For MM-NN, we show the results reported in the original paper, MAP scores at some code length are missing.}
and SePH$_{km}$ are nonlinear modeling methods, and others are linear modeling methods.
The code length is set to be 16, 32, 64, and 128 bits, respectively.

\begin{table}
\begin{center}
\begin{tabular}{|c|c|c|c|c|c|c|}
\hline
\multirow{2}{*}{Tasks}      &\multirow{2}{*}{Methods}    &\multicolumn{4}{c|}{Code Length}\\
\cline{3-6}
                              &        & $K$=16           & $K$=32      & $K$=64     & $K$=128\\
\hline\hline
\multirow{8}{*}{\tabincell{c}{Image query \\ vs. \\ Text database}}
                        &CVH       &   0.3687                     & 0.4182             & 0.4602       & 0.4466         \\ \cline{2-6}
                         &IMH       &    0.4187                  & 0.3975           & 0.3778         &  0.3668        \\   \cline{2-6}
                       &LSSH        &   0.3900                  & 0.3924            & 0.3962       &  0.3966        \\   \cline{2-6}
                       &CMFH       &    0.4267                  & 0.4229           & 0.4207       &   0.4182        \\   \cline{2-6}
                        &CMSSH        &   0.4063                   & 0.3927        & 0.3939          & 0.3739         \\     \cline{2-6}
                        &SCM-Seq      &    0.4842                & 0.4941         & 0.4947          &  0.4965      \\  \cline{2-6}
                         &SePH$_{km}$      &    0.5421                 & 0.5499           & 0.5537          & 0.5601       \\  \cline{2-6}
                         &MM-NN     &    57.44                 & --           & 56.33          & --       \\  \cline{2-6}
\hline\hline
\multirow{8}{*}{\tabincell{c}{Text query \\ vs. \\ Image database}}
                        &CVH       &   0.3646                   & 0.4024           & 0.4339         &  0.4255        \\ \cline{2-6}
                        &IMH       &    0.4053                  & 0.3892            & 0.3758         &   0.3627        \\   \cline{2-6}
                       &LSSH        &    0.4286                  & 0.4248           & 0.4248         & 0.4175          \\   \cline{2-6}
                       &CMFH       &    0.4627                  & 0.4556             & 0.4518        &  0.4478         \\   \cline{2-6}
                        &CMSSH        &  0.3874                   & 0.3849        & 0.3704          &  0.3699       \\     \cline{2-6}
                        &SCM-Seq      &    0.4536                 & 0.4620          & 0.4630          &  0.4644    \\  \cline{2-6}
                         &SePH$_{km}$      &    0.6302                & 0.6425          & 0.6506          &  0.6580      \\  \cline{2-6}
                         &MM-NN      &    56.91                & --          & 55.83          &  --      \\  \cline{2-6}
\hline
\end{tabular}
\end{center}
\caption{MAP comparison of different cross-modal hashing methods on the NUS-WIDE dataset.}
\label{tab:mapNUSHash}
\end{table}

From the experimental results, we can draw the following observations.

1) IMH performs better than CVH. The reason is that CVH only exploits the inter-modality similarity,
but IMH exploits both inter-modality and intra-modality similarity.
So it is useful to model the intra-modal similarity in the cross-modal hashing algorithms.

2) CMSSH performs better than CVH. The reason is that CVH only uses the pairwise information,
but CMSSH uses both similar and dissimilar pairs.
So using more information is helpful for improving performance.

3) Although without supervised information, experiments showed that
CMFH and LSSH can well exploit the latent semantic affinities of training data
and yield state-of-the-art performance for cross-modal retrieval.
So modeling the common representation space in a appropriate manner (like CMFH and LSSH) is very important.

4) SCM-Seq integrates semantic labels into the hashing learning procedure via maximizing semantic correlations,
which outperforms several state-of-the-art methods.
It can be seen that supervised information is beneficial for learning binary codes for various modalities of data.

5) SePH performs better than other methods, due to its capability to
better preserve semantic affinities in Hamming space,
as well as the effectiveness of kernel logistic regression
to model the non-linear projections from features to hash codes.
MM-NN learns non-linear projection by using the siamese neural network architecture.
It generally performs better than the linear models.
So learning non-linear projection is more appropriate for complex structure of multimodal data.

6) As the length of hash codes increases, the performance of supervised hashing methods (SCM-Seq and SePH) keeps increasing,
which reflects the capability of utilizing longer hash codes to better preserve semantic affinities.
Meanwhile, performance of some baselines like CVH, IMH, CMSSH decreases,
which is also observed in previous work \cite{ding2014collective,zhang2014large,Zhu2013LCMH}.
So it is more difficult to learn longer binary codes without supervised information.

\section{Discussion and future trends}

Although some promising results have been achieved in the
field of cross-modal retrieval, there is still a gap between state-of-the-art methods and
user expectation, which indicates that we still need to investigate the cross-modal retrieval problem.
In the following, we discuss the future research opportunities for cross-modal retrieval.

1. Collection of multimodal large-scale datasets

Now researchers have been working hard to design more and more sophisticated algorithms to retrieve or summarize the multimodal data.
However, there is a lack of good sources for further training, testing, evaluating and comparing the performance of different algorithms.
Currently available datasets for cross-modal retrieval research are either too small such as the Wikipedia dataset that only contains 2866 documents,
or too specific such as NUS-WIDE only consists of user tags.
Hence, it would be tremendously helpful for researchers if there exists a multimodal large-scale dataset,
which contains more than two modalities of data and large-scale multimodal data with ground truth.

2. Multimodal learning with limited and noisy annotations

The emerging applications on social networks have produced huge amount of
multimodal content created by people.
Typical examples include Flickr, YouTube, Facebook, MySpace, WeiBo, WeiXin, etc.
It is well known that the multimodal data in the web is loosely organized,
and the annotations of these data are limited and noisy.
Obviously, it is difficult to label large scale multimodal data.
However, the annotations provide semantic information for multimodal data,
so how to utilize the limited and noisy annotations to learn semantic correlations among the multimodal data
in this scenario need to be addressed in the future.

3. Scalability on large-scale data

Driven by wide availability of massive storage devices, mobile devices
and fast networks, more and more multimedia resources are generated and propagated on
the web. With rapid growth of the multimodal data, we need develop effective and efficient
algorithms that are scalable to distributed platforms.
We also need to conduct further research on effectively and efficiently organizing
each relevant modality of data together.

4. Deep learning on multimodal data

Recently, deep learning algorithms achieve much progress in image classification \cite{Alex2012Net,Simo2015VGG,Szegedy2015Googlenet},
video recognition/classsification \cite{Karpathy14Video,Simonyan14TwoStream}, text analysis \cite{mao2015deep,Oriol15CaptionGenerator}, and so on.
Deep learning algorithms show good properties in representation learning.
Powerful representations are helpful for reducing heterogeneity gap and semantic gap between different modalities of data.
Hence, combining appropriate deep learning algorithms to model different types of data for cross-modal retrieval
(such as CNN for modeling images, RNN (Recurrent Neural Networks) for modeling text) is a future trend.

5. Finer-level cross-modal semantic correlation modeling

 Most of published works usually embed different modalities into a common embedding space.
 For example, they map images and texts into a common space, where different modalities of data can be compared.
 However, this is too rough because different image fragments correspond to different text fragments,
 and considering such finer correspondence could explore the image-text semantic relations more accurately.
 So how to obtain the fragments of different modalities and find their correspondence are very important.
 Accordingly, new model should be designed for modeling such complex relations.

\section{Conclusions}

Cross-modal retrieval provides an effective and powerful way to multimodal data retrieval
and it is more convenient than traditional single-modality-based techniques.
This paper gives an overview of cross-modal retrieval,  summarizes a number of representative methods
and classifies them into two main groups:
1) real-valued representation learning, and 2) bianry representation learning.
Then, we introduce several commonly used multimodal datasets, and
empirically evaluate the performance of some representative methods on some commonly used datasets.
We also discuss the future trends in cross-modal retrieval field.
Although significant work has been carried out in this field,
cross-modal retrieval has not been well-addressed to date.
There is still much work to be done to better process
cross-modal retrieval.
We expect this paper will help readers to understand the state-of-the-art in cross-modal retrieval and motivate more meaningful works.

\bibliographystyle{IEEEtran}
\bibliography{reference}
\end{document}